\documentclass[a4paper,USenglish,cleveref, autoref, thm-restate, 11pt]{lipics-v2021-custom-font}
\usepackage{placeins}
\usepackage{tikz}
\usetikzlibrary{calc}
\usetikzlibrary{patterns}
\usetikzlibrary{decorations.pathreplacing}
\usetikzlibrary{fadings}
\usepackage[linesnumbered,ruled,vlined]{algorithm2e}
\title{The Limits of Local Search for the Maximum Weight Independent Set Problem in \texorpdfstring{$d$}{d}-Claw Free Graphs} %TODO Please add

\titlerunning{The Limits of Local Search for the MWIS in \texorpdfstring{$d$}{d}-Claw Free Graphs} %TODO optional, please use if title is longer than one line

\author{Meike Neuwohner}{Forschungsinstitut f\"ur Diskrete Mathematik, Universit\"at Bonn}{neuwohner@or.uni-bonn.de}{https://orcid.org/0000-0002-3664-3687}{}%TODO mandatory, please use full name; only 1 author per \author macro; first two parameters are mandatory, other parameters can be empty. Please provide at least the name of the affiliation and the country. The full address is optional

\authorrunning{M. Neuwohner} %TODO mandatory. First: Use abbreviated first/middle names. Second (only in severe cases): Use first author plus 'et al.'

\Copyright{Meike Neuwohner} %TODO mandatory, please use full first names. LIPIcs license is "CC-BY";  http://creativecommons.org/licenses/by/3.0/

\ccsdesc[500]{Theory of computation~Packing and covering problems} %TODO mandatory: Please choose ACM 2012 classifications from https://dl.acm.org/ccs/ccs_flat.cfm 

\keywords{\texorpdfstring{$d$}{d}-claw free graphs, independent set, local search, weighted \texorpdfstring{$k$}{k}-Set Packing} %TODO mandatory; please add comma-separated list of keywords

\relatedversion{} %optional, e.g. full version hosted on arXiv, HAL, or other respository/website
\supplement{}%optional, e.g. related research data, source code, ... hosted on a repository like zenodo, figshare, GitHub, ...

\nolinenumbers %uncomment to disable line numbering

\hideLIPIcs  %uncomment to remove references to LIPIcs series (logo, DOI, ...), e.g. when preparing a pre-final version to be uploaded to arXiv or another public repository

\EventEditors{John Q. Open and Joan R. Access}
\EventNoEds{2}
\EventLongTitle{42nd Conference on Very Important Topics (CVIT 2016)}
\EventShortTitle{CVIT 2016}
\EventAcronym{CVIT}
\EventYear{2016}
\EventDate{December 24--27, 2016}
\EventLocation{Little Whinging, United Kingdom}
\EventLogo{}
\SeriesVolume{42}
\ArticleNo{23}
\newtheorem{notation}[theorem]{Notation}
\newtheorem*{RestateTheoMainResult}{Theorem~\ref{TheoMainResult}}
        \newcommand*{\chrg}[2]{\mathrm{charge}(#1,#2)}
        
        \newcommand*{\contr}[2]{\mathrm{contr}(#1,#2)}
\begin{document}

\maketitle

%TODO mandatory: add short abstract of the document
\begin{abstract}
We consider the Maximum Weight Independent Set Problem (MWIS) in $d$-claw free graphs, i.e.\ the task of computing an independent set of maximum weight in a given $d$-claw free graph $G=(V,E)$ equipped with a positive weight function $w:V\rightarrow\mathbb{R}_{>0}$. For $k\geq 1$, the MWIS in $k+1$-claw free graphs generalizes the weighted $k$-Set Packing Problem, where one is given a collection $\mathcal{S}$ of sets, each of cardinality at most $k$, and a positive weight function $w:\mathcal{S}\rightarrow\mathbb{R}_{>0}$, and has to find a sub-collection of $\mathcal{S}$ consisting of pairwise disjoint sets of maximum total weight. Given that for $k\geq 3$, this problem does not permit a polynomial time $o(\frac{k}{\log k})$-approximation unless $P=NP$ \cite{LowerBoundKSetPacking}, most previous algorithms for both weighted $k$-Set Packing and the MWIS in $d$-claw free graphs rely on local search. For the last twenty years, the algorithm \emph{SquareImp} proposed by Berman \cite{Berman} for the MWIS in $d$-claw free graphs has remained unchallenged for both problems. It searches for a certain type of local improvement of the squared weight function, and can achieve a performance ratio of $\frac{d}{2}+\epsilon$ for for any fixed $\epsilon>0$ (implying a $\frac{k+1}{2}+\epsilon$-approximation for weighted $k$-Set Packing). Recently, Berman's algorithm was improved by Neuwohner \cite{NeuwohnerLipics}, obtaining an approximation guarantee slightly below $\frac{d}{2}$, and inevitably raising the question of how far one can get by using local search.\\ In this paper, we finally answer this question asymptotically in the following sense:  By considering local improvements of logarithmic size, we obtain approximation ratios of $\frac{d-1+\epsilon_d}{2}$ for the MWIS in $d$-claw free graphs for $d\geq 3$ in quasi-polynomial time, where $0\leq \epsilon_d\leq 1$ and $\lim_{d\rightarrow\infty}\epsilon_d = 0$. By employing the color coding technique, we can use the previous result to obtain a polynomial time $\frac{k+\epsilon_{k+1}}{2}$-approximation for weighted $k$-Set Packing. On the other hand, we provide examples showing that no local improvement algorithm considering local improvements of size $\mathcal{O}(\log(|\mathcal{S}|))$ with respect to some power $w^\alpha$ of the weight function, where $\alpha\in\mathbb{R}$ is chosen arbitrarily, but fixed, can yield an approximation guarantee better than $\frac{k}{2}$ for the weighted $k$-Set Packing Problem with $k\geq 3$.
\end{abstract}
\newpage
 \section{Introduction}
\begin{figure}[t]
  \centering
		\begin{tikzpicture}[scale = 0.5,elem/.style = {circle, draw = black, thick, fill = none, inner sep = 0.5mm, minimum size = 4mm}]
		\node[elem] (C) at (0,0){$v_0$};
		\node[elem] (A1) at (3,2){$v_1$};
		\node[elem] (A2) at (3,0){$v_2$};
		\node[elem] (A3) at (3,-2){$v_3$};
		\draw (C)--(A1);
		\draw (C)--(A2);
		\draw (C)--(A3);
		\draw[ decoration = brace, decorate] (3.7,2.5) to (3.7,-2.5);
		\node at (-1.2,1.2) { center node};
		\node[align = left] at (5.5,0){ $d$ talons\\
		$\rightsquigarrow T_C$};
		\end{tikzpicture}
		\caption{A $d$-claw $C$ for $d=3$.}\label{FigClaw}
	\end{figure}
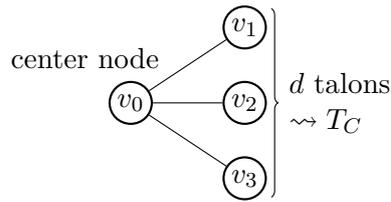

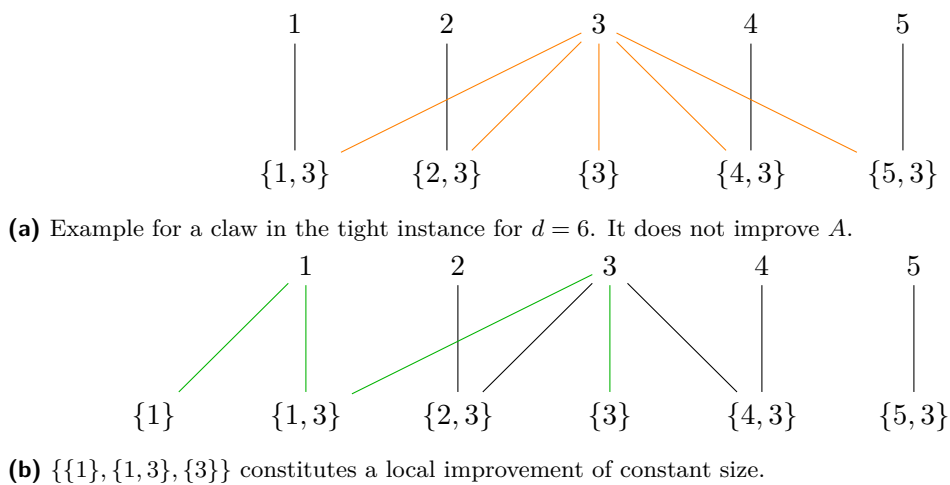
\begin{figure}[t]
 \centering
 \begin{subfigure}{\textwidth}
 \centering
	\begin{tikzpicture}
	\node (11) at (-2,0){};
	\node (1) at (0,2) {$1$};
	\node (2) at (2,2) {$2$};
	\node (3) at (4,2) {$3$};
	\node (4) at (6,2) {$4$};
	\node (5) at (8,2) {$5$};
	\node (13) at (0,0) {$\{1,3\}$};
	\node (23) at (2,0) {$\{2,3\}$};
	\node (33) at (4,0) {$\{3\}$};
	\node (43) at (6,0) {$\{4,3\}$};
	\node (53) at (8,0) {$\{5,3\}$};
	\draw (1)--(13);
	\draw (2)--(23);
	\draw[orange] (3)--(33);
	\draw (4)--(43);
	\draw (5)--(53);
	\draw[orange] (3)--(13);
	\draw[orange] (3)--(23);
	\draw[orange] (3)--(43);
	\draw[orange] (3)--(53);
	\end{tikzpicture}\caption{Example for a claw in the tight instance for $d=6$. It does not improve $A$.}\end{subfigure}
	\begin{subfigure}{\textwidth}
	\centering\begin{tikzpicture}
	\node (1) at (0,2){$1$};
	\node (11) at (-2,0) {$\{1\}$};
	\node (2) at (2,2) {$2$};
	\node (3) at (4,2) {$3$};
	\node (4) at (6,2) {$4$};
	\node (5) at (8,2) {$5$};
	\node (13) at (0,0) {$\{1,3\}$};
	\node (23) at (2,0) {$\{2,3\}$};
	\node (33) at (4,0) {$\{3\}$};
	\node (43) at (6,0) {$\{4,3\}$};
	\node (53) at (8,0) {$\{5,3\}$};
	\draw[green!70!black] (1)--(13);
	\draw (2)--(23);
	\draw[green!70!black] (3)--(33);
	\draw (4)--(43);
	\draw (5)--(53);
	\draw[green!70!black] (3)--(13);
	\draw (3)--(23);
	\draw (3)--(43);
	\draw[green!70!black] (1) --(11);
	\end{tikzpicture}\caption{$\{\{1\},\{1,3\},\{3\}\}$ constitutes a local improvement of constant size.}\end{subfigure}
	\caption{(Part of) the tight instance provided in \cite{Berman}.}\label{FigTightExample}
\end{figure}

 For $d\geq 1$, a \emph{$d$-claw} $C$ \cite{Berman} is defined to be a star consisting of one \emph{center node} and a set $T_C$ of $d$ additional vertices connected to it, which are called the \emph{talons} of the claw (see Figure \ref{FigClaw}). Moreover, similar to \cite{Berman}, we define a $0$-claw to be a graph consisting only of a single vertex $v$, which is regarded as the unique element of $T_C$ in this case. An undirected graph $G=(V,E)$ is said to be \emph{$d$-claw free} if none of its induced subgraphs forms a $d$-claw. For example, $1$-claw free graphs do not possess any edges, while $2$-claw free graphs are disjoint unions of cliques.
 For natural numbers $k\geq 3$, the Maximum Weight Independent Set Problem (MWIS) in $k+1$-claw free graphs is often studied as a generalization of the weighted $k$-Set Packing Problem. The latter is defined as follows: Given a family $\mathcal{S}$ of sets each of size at most $k$ together with a positive weight function $w:\mathcal{S}\rightarrow\mathbb{R}_{>0}$, the task is to find a sub-collection $A$ of $\mathcal{S}$ of maximum weight such that the sets in $A$ are pairwise disjoint. By considering the \emph{conflict graph} $G_\mathcal{S}$ associated with an instance of the weighted $k$-Set Packing Problem, one obtains a weight preserving one-to-one correspondence between feasible solutions to the $k$-Set Packing Problem and independent sets in $G_\mathcal{S}$. The vertices of $G_\mathcal{S}$ are given by the sets in $\mathcal{S}$ and the edges represent non-empty set intersections. It is not hard to see that $G_\mathcal{S}$ is $k+1$-claw free.\\
 As far as the cardinality version of the $k$-Set Packing Problem is concerned, considerable progress has been made over the last decade. The first improvement upon the approximation guarantee of $k$ achieved by a simple greedy approach was obtained by Hurkens and Schrijver in 1989 \cite{HurkensSchrijver}, who showed that for any $\epsilon>0$, there exists a constant $p_\epsilon$ for which a local improvement algorithm that first computes a maximal collection of disjoint sets and then repeatedly applies local improvements of constant size at most $p_\epsilon$, until no more exist, yields an approximation guarantee of $\frac{k}{2}+\epsilon$. In this context, a disjoint collection $X$ of sets contained in the complement of the current solution $A$ is considered a \emph{local improvement of size $|X|$} if the sets in $X$ intersect at most $|X|-1$ sets from $A$, which are then replaced by the sets in $X$, increasing the cardinality of the found solution. Hurkens and Schrijver also proved that a performance guarantee of $\frac{k}{2}$ is best possible for a local search algorithm only considering improvements of constant size, while Hazan, Safra and Schwartz \cite{LowerBoundKSetPacking} established in 2006 that no $o(\frac{k}{\log k})$-approximation algorithm is possible in general unless $P=NP$. At the cost of a quasi-polynomial runtime, Halld\'orsson \cite{Halldorsson} proved an approximation factor of $\frac{k+2}{3}$ by applying local improvements of size logarithmic in the total number of sets. Cygan, Grandoni and Mastrolilli \cite{CyganGrandoniMastrolilli} managed to get down to an approximation factor of $\frac{k+1}{3}+\epsilon$, still with a quasi-polynomial runtime.\\ The first polynomial time algorithm improving on the result by Hurkens and Schrijver was obtained by Sviridenko and Ward \cite{SviridenkoWard} in 2013. By combining means of color coding with the algorithm presented in \cite{Halldorsson}, they achieved an approximation ratio of $\frac{k+2}{3}$. This result was further improved to $\frac{k+1}{3}+\epsilon$ for any fixed $\epsilon>0$ by Cygan \cite{Cygan}, obtaining a polynomial runtime doubly exponential in $\frac{1}{\epsilon}$. The best approximation algorithm for the unweighted $k$-Set Packing Problem in terms of performance ratio and running time is due to F\"urer and Yu from 2014 \cite{FurerYu}, who achieved the same approximation guarantee as Cygan, but a runtime only singly exponential in $\frac{1}{\epsilon}$.\\
 Concerning the unweighted version of the MWIS in $d$-claw free graphs, as remarked in \cite{SviridenkoWard}, both the result of Hurkens and Schrijver as well as the quasi-polynomial time algorithms by Halld\'orsson and Cygan, Grandoni and Mastrolilli  translate to this more general context, yielding approximation guarantees of $\frac{d-1}{2}+\epsilon$, $\frac{d+1}{3}$ and $\frac{d}{3}+\epsilon$, respectively. However, it is not clear how to extend the color coding approach relying on coloring the underlying universe to the setting of $d$-claw free graphs \cite{SviridenkoWard}.\\
 When it comes to the weighted variant of the problem, even less is known.  For $d\leq 3$, it is solvable in polynomial time (see \cite{Minty} and \cite{sbihi1980algorithme} for the unweighted, \cite{nakamura2001revision} for the weighted variant), while for $d\geq 4$, again no $o(\frac{d}{\log d})$-approximation algorithm is possible unless $P=NP$ \cite{LowerBoundKSetPacking}. Moreover, in contrast to the unit weight case, considering local improvements of constant size can only slightly improve on the performance ratio of $d-1$ obtained by the greedy algorithm since Arkin and Hassin have shown that such an approach yields an approximation ratio no better than $d-2$ in general \cite{ArkinHassin}. Here, analogously to the unweighted case, given an independent set $A$, an independent set $X$ is called a \emph{local improvement of $A$} if it is disjoint from $A$ and the total weight of the neighbors of $X$ in $A$ is strictly smaller than the weight of $X$. Despite the negative result in \cite{ArkinHassin}, Chandra and Halld\'orsson \cite{ChandraHalldorsson} have found that if one does not perform the local improvements in an arbitrary order, but in each step augments the current solution $A$ by an improvement $X$ that maximizes the ratio between the total weight of the vertices added to and removed from $A$ (if exists), the resulting algorithm, which the authors call \emph{BestImp}, approximates the optimum solution within a factor of $\frac{2d}{3}$. By scaling and truncating the weight function to ensure a polynomial number of iterations, they obtain a $\frac{2d}{3}+\epsilon$-approximation algorithm for the MWIS in $d$-claw free graphs.\\
 For 20 years, the algorithm \emph{SquareImp} devised by Berman \cite{Berman} has been the state-of-the-art for both weighted $k$-Set Packing and the MWIS in $d$-claw free graphs. SquareImp iteratively applies local improvements of the squared weight function that arise as sets of talons of claws in $G$, until no more exist. In doing so, SquareImp achieves an approximation ratio of $\frac{d}{2}$, leading to a polynomial time $\frac{d}{2}+\epsilon$-approximation algorithm for the MWIS in $d$-claw free graphs for any fixed $\epsilon >0$.\\ Berman also provides an example for $w\equiv 1$ showing that his analysis is tight. It consists of a bipartite graph $G=(V,E)$ the vertex set of which splits into a maximal independent set $A=\{1,\dots,d-1\}$ such that no claw improves $|A|$, and an optimum solution $B=\binom{A}{1}\cup\binom{A}{2}$, where the set of edges is given by $E=\{\{a,b\}:a\in A, b\in B, a\in b\}$. As the example uses unit weights, he also concludes that applying the same type of local improvement algorithm for a different power of the weight function does not provide further improvements.\\ However, as also implied by the result in \cite{HurkensSchrijver}, while no small improvements \emph{forming the set of talons of a claw} in the input graph exist in the tight example given by Berman, once this additional condition is dropped, improvements of small constant size can be found quite easily (see Figure~\ref{FigTightExample}). This observation is the basis of a recent paper by Neuwohner \cite{NeuwohnerLipics}, who managed to obtain an approximation guarantee slightly below $\frac{d}{2}$ by taking into account a broader class of local improvements, namely all improvements of the squared weight function of size at most $(d-1)^2+(d-1)$.

 In this paper, following \cite{NeuwohnerLipics}, we revisit the analysis of the algorithm SquareImp proposed by Berman and show that whenever it is close to being tight, the instance is locally unweighted in the sense that almost every time when a vertex from the solution chosen by SquareImp and a vertex from any optimum solution share an edge, their weights must be very similar. However, while \cite{NeuwohnerLipics} merely focuses on one of the two major steps in Berman's analysis, we consider both of them, allowing us to derive much stronger statements concerning the structure of instances where SquareImp does not do much better than a $\frac{d}{2}$-approximation. In particular, we are able to transfer techniques that are used in the state-of-the-art works on the unweighted $k$-Set Packing Problems (cf. \cite{SviridenkoWard}, \cite{FurerYu}) to a setting where vertex weights are locally similar. This is the main ingredient for our algorithm \emph{LogImp}, which, in addition to the type of improvements considered by SquareImp, searches for a certain type of local improvement of logarithmic size. In doing so, it obtains an approximation guarantee of $\frac{d-1+\epsilon_d}{2}$ for the MWIS in $d$-claw free graphs for $d\geq 3$, where $0\leq \epsilon_d \leq 1$ and $\lim_{d\rightarrow\infty} \epsilon_d = 0$.\\  While we can only guarantee a quasi-polynomial running time for the MWIS in $d$-claw free graphs, we manage to obtain a polynomial time $\frac{k+\epsilon_{k+1}}{2}$-approximation algorithm for the weighted $k$-Set Packing Problem by means of color coding.\\ Furthermore, we provide examples showing that any local improvement algorithm that, for an arbitrarily chosen, but fixed parameter $\alpha\in\mathbb{R}$, searches for local improvements of $w^\alpha$ of size $\mathcal{O}(\log(|\mathcal{S}|))$, cannot produce an approximation guarantee better than $\frac{k}{2}$ for the weighted $k$-Set Packing Problem with $k\geq 3$. Note that this also implies an analogous statement for the more general MWIS in $d$-claw free graphs, substituting $k$ by $d-1$. \\
 \section{Our Contribution}
 In the following, we provide an outline of our results, giving some more details concerning the main ideas and techniques employed. As indicated in the abstract, our main contribution is to present a local improvement algorithm for the weighted $k$-Set Packing Problem which is asymptotically best possible in the sense that for $k\rightarrow\infty$, the absolute difference between the approximation guarantee of our algorithm LogImp and the best that is possible via pure local search tends to zero. This result is interesting for several reasons:\\
 First, the recent improvement over an approximation guarantee of $\frac{d}{2}$ by Neuwohner~\cite{NeuwohnerLipics} has once again raised the question how far one can push the approximation guarantees for the weighted $k$-Set Packing Problem using local search, which has now finally been answered. In doing so, we are the first ones to consider local improvements of logarithmic size and apply more advanced techniques such as color coding to the weighted setting of the problem. While these methods have been the state-of-the-art for the unweighted problem for several years now \cite{Cygan},\cite{FurerYu}, only very simple (to describe) algorithms searching for local improvements of constant size have been considered for the weighted case so far.\\
 Furthermore, our lower bound example significantly extends the current state of knowledge in that the one lower bound example specifically tailored to the weighted case \cite{ArkinHassin} only considers local improvements of constant size and the original weight function. Additionally, our lower bound construction employs results by Erd\H{o}s and Sachs \cite{ErdosSachs} on the existence of regular graphs of logarithmic girth, for which we derive upper bounds on density (i.e.\ the ratio between the number of edges and the number of vertices) in subgraphs induced by a logarithmic number of nodes.\\
 Another striking point is the fact that different from the unit weight case where the step from improvements of constant to improvements of logarithmic size can improve the obtainable approximation guarantee by a factor of roughly $\frac{2}{3}$ \cite{HurkensSchrijver},\cite{CyganGrandoniMastrolilli},\cite{FurerYu}, in the weighted setting, the approximation guarantees that can be achieved by considering local improvements of constant respectively logarithmic size are asymptotically the same in that their ratio converges to $1$ for $d\rightarrow\infty$.\\
 Further contributions of our paper lie in the methods that we employ in our analysis. In order to prove the existence of local improvements of logarithmic size, the state-of-the-art work concerning the unit weight variant of the $k$-Set Packing Problem \cite{FurerYu} relies on a result originally proven by Berman and F\"urer \cite{berman1994approximating}. It states that any graph for which the number of edges is by a constant factor larger than the number of vertices, contains a cycle and a subgraph with more edges than vertices of logarithmic size. In Section~\ref{SecLogSizeImprovements}, we port this idea to a weighted setting with locally similar weights. More precisely, we show that given a graph $G=(V,E)$, equipped with strictly positive vertex weights such that the ratio between the weights of adjacent vertices is close to $1$ and with the property that all vertices in $V$ except for a subset $Y$ of small total weight have a large degree, we can find both a cycle in $G$ and a subgraph of $G$ with more edges than vertices of logarithmic size. Note that this is not a direct consequence of the unweighted statement because although the weight of $Y$ is small, its cardinality can still be large compared to $|V|$.\\
 Equipped with this result, we study an algorithm that we call LogImp, which searches for two types of local improvements with respect to the squared weight function. The first type of local improvements are those also considered by Berman's algorithm SquareImp, which we call \emph{claw-shaped}. Cycles of logarithmic size in an auxiliary graph constitute the backbone of the \emph{circular improvements}, the second type of local improvement we consider.\\
  To analyze our algorithm LogImp, we fully classify the vertices in an optimum solution $A^*$ in relation to any independent set $A$ for which no claw-shaped or circular improvement exists. In doing so, we observe that each $v\in A$ with less than $d-1$ neighbors (a.k.a.\ more than one \emph{missing} neighbor) in $A^*$ improves our bound on $w(A^*)$ by a constant fraction of its weight. Moreover, we show that for each $v\in A$, each of its neighbors in $A^*$ falls into one of two categories, which we denote by \emph{profitable} and \emph{helpful}. While each profitable neighbor of $v$ improves the bound $w(A^*)\leq\frac{d}{2}\cdot w(A)$ we obtain by applying Berman's analysis \cite{Berman} to our algorithm by a constant fraction of $w(v)$, each helpful vertex either contributes a significant amount towards a claw-shaped improvement, or induces an edge in an auxiliary graph. Under certain additional assumptions, cycles of logarithmic size in this auxiliary graph yield circular improvements. Roughly speaking, the latter implies an upper bound on the number of helpful vertices and, hence, a lower bound on the number of missing or profitable vertices, resulting in an improved approximation guarantee.\\
 The rest of this paper is organized as follows:
 In Section~\ref{SecPrelim}, we review the algorithm SquareImp by Berman and give a short overview of the analysis pointing out the results we reuse in the analysis of our algorithm. In the next section, we prove the existence of local improvements of logarithmic size in a weighted setting with locally similar weights. In Section~\ref{SecAlgo}, we introduce our algorithm LogImp, (which stands for \emph{log}arithmic \emph{imp}rovement,) and we prove the first part of our main result, which is given by Theorem~\ref{TheoSequenceEpsilons}. The second part, i.e.\ how to achieve a polynomial running time for the weighted $k$-Set Packing Problem, is discussed in Section~\ref{SecPolyTime}.
 \begin{theorem}
 There exists a sequence $(\epsilon_d)_{d\geq 3}\in[0,1]^{\mathbb{N}_{\geq 3}}$ with $\lim_{d\rightarrow\infty}\epsilon_d =0$ and, for each $d\geq 3$, an algorithm for the MWIS in $d$-claw free graphs that runs in quasi-polynomial time (considering $d$ a constant) and achieves an approximation guarantee of $\frac{d-1+\epsilon_d}{2}$.\\
  On instances that arise as the conflict graph of a (known) instance of the weighted $d-1$-Set Packing Problem, the algorithm can be implemented to run in polynomial time.\label{TheoSequenceEpsilons}	
 \end{theorem} Section~\ref{SecExamples} shows that any local improvement algorithm that, for an arbitrarily chosen, but fixed parameter $\alpha\in\mathbb{R}$, searches for local improvements of $w^\alpha$ of size $\mathcal{O}(\log(|\mathcal{S}|))$, cannot produce an approximation guarantee better than $\frac{k}{2}$ for the weighted $k$-Set Packing Problem with $k\geq 3$.  Finally, Section~\ref{SecConclusion} provides some concluding remarks.

 %TODO say that we take notation of claws from Berman
 \section{Preliminaries\label{SecPrelim}}
 In this section, we shortly recap the definitions and main results from \cite{Berman} that we will employ in the analysis of our local improvement algorithm. We first introduce some basic notation that is needed for its formal description.
 \begin{definition}[Neighborhood \cite{Berman}]
Given an undirected graph $G=(V,E)$ and subsets $U,W\subseteq V$ of its vertices, we define the \emph{neighborhood} $N(U,W)$ of $U$ in $W$ as 
\[N(U,W):=\{w\in W:\exists u\in U: \{u,w\}\in E \vee u=w\}.\]
In order to simplify notation, for $u\in V$ and $W\subseteq V$, we write $N(u,W)$ instead of $N(\{u\}, W)$.\label{DefNeighborhood}
\end{definition}
\begin{notation}
Given a weight function $w:V\rightarrow\mathbb{R}$ and some $U\subseteq V$, we write \linebreak[4]$w^2(U):=\sum_{u\in U} w^2(u)$. Observe that in general, $w^2(U)\neq (w(U))^2$.
\end{notation}
\begin{definition}[\cite{Berman}]
Given an undirected graph $G=(V,E)$, a weight function $w:V\rightarrow\mathbb{R}_{\geq 0}$ and an independent set $A\subseteq V$, we say that a vertex set $B\subseteq V$ \emph{improves $w^2(A)$} if $B$ is independent in $G$ and $w^2(A\backslash N(B,A)\cup B) > w^2(A)$ holds.
For a claw $C$ in $G$, we say that $C$ improves $w^2(A)$ if its set of talons $T_C$ does.	
\end{definition}
Note that an independent set $B$ improves $A$ if and only if we have \mbox{$w^2(B)>w^2(N(B,A))$} (see Proposition~\ref{PropEquivDefLocalImpr}).\\ Using the notation introduced above, Berman's algorithm SquareImp \cite{Berman} can now be formulated as in Algorithm~\ref{AlgoSquareImp}.
\begin{algorithm}[t]
	\DontPrintSemicolon
\KwIn{an undirected $d$-claw free graph $G=(V,E)$ and a (positive) weight function $w:V\rightarrow\mathbb{R}_{>0}$\;}
\KwOut{an independent set $A\subseteq V$\;}
$A\gets \emptyset$\;
\While{there exists a claw $C$ in $G$ that improves $w^2(A)$}
{$A\gets A\backslash N(T_C, A)\cup T_C$\;}
\Return $A$\;
\caption{SquareImp \cite{Berman}}\label{AlgoSquareImp}	
\end{algorithm}
Observe that by positivity of the weight function, every $v\not\in A$ such that $A\cup\{v\}$ is independent constitutes the talon of a $0$-claw improving $w^2(A)$, so the algorithm returns a maximal independent set.\\
The main idea of the analysis of SquareImp presented in \cite{Berman} is to charge the vertices in $A$ for preventing adjacent vertices in an optimum solution $A^*$ from being included into $A$. The latter is done by spreading the weight of the vertices in $A^*$ among their neighbors in the maximal independent set $A$ in such a way that no vertex in $A$ receives more than $\frac{d}{2}$ times its own weight. The suggested distribution of weights proceeds in two steps:\\ First, each vertex $u\in A^*$ invokes costs of $\frac{w(v)}{2}$ at each $v\in N(u,A)$, leaving a remaining weight of $w(u)-\frac{w(N(u,A))}{2}$ to be distributed. (Note that this term can be negative.)\\ In a second step, each vertex in $u$ sends an amount of $w(u)-\frac{w(N(u,A))}{2}$ to a heaviest neighbor it possesses in $A$, which is captured by the following definition of \emph{charges}:
\begin{definition}[Charges \cite{Berman}]
Let $G=(V,E)$ be an undirected graph and let \mbox{$w:V\rightarrow\mathbb{R}_{>0}$} be a (positive) weight function. Further assume that an independent set $A^*\subseteq V$ and a maximal independent set $A\subseteq V$ are given. We define a map $\mathrm{charge}:A^*\times A\rightarrow \mathbb{R}$ as follows:\\
For each $u\in A^*$, pick a vertex $v\in N(u,A)$ of maximum weight and call it $n(u)$. Observe that this is possible, because $A$ is a maximal independent set in $G$, implying that $N(u,A)\neq \emptyset$ since either $u\in A$ or $u$ possesses a neighbor in $A$.\\
Next, for $u\in A^*$ and $v\in A$, define \[\chrg{u}{v}:=\begin{cases}
w(u)-\frac{1}{2}w(N(u,A)) &,\text{ if }v=n(u)\\
0&,\text{ otherwise. }
\end{cases}\] \label{DefCharges}
\end{definition}
The definition of charges directly implies the subsequent statement:
\begin{corollary}[\cite{Berman}]
 In the situation of Definition~\ref{DefCharges}, we have
 \begin{align*}w(A^*)&=\sum_{u\in A^*} \frac{w(N(u,A))}{2}+\sum_{u\in A^*} \chrg{u}{n(u)}\\&\leq \sum_{u\in A^*} \frac{w(N(u,A))}{2}+\sum_{u\in A^*: \chrg{u}{n(u)}>0} \chrg{u}{n(u)}.\end{align*} \label{CorBoundwAstar}
\end{corollary}

The analysis proposed by Berman now proceeds by bounding the total weight  sent to the vertices in $A$ during the two steps of the cost distribution separately. In doing so, Lemma~\ref{LemWeightsNeighborhoodsdminus1} bounds the weight received in the first step, while Lemma~\ref{LemPropPositiveCharges} and Lemma~\ref{LemBoundCharges}  take care of the total charges invoked. The following results appear in \cite{Berman}, but we have slightly modified the way they are formulated to suit our purposes. Matching proofs, which are partly easier than those presented in \cite{Berman}, can be found in Appendix~\ref{AppendixAnalysisSquareImp}.
\begin{lemma}[\cite{Berman}]
 In the situation of Definition~\ref{DefCharges}, if the graph $G$ is $d$-claw free for some $d\geq 2$, then
 \[\sum_{u\in A^*} \frac{w(N(u,A))}{2}\leq \frac{d-1}{2}\cdot w(A).\] \label{LemWeightsNeighborhoodsdminus1}
\end{lemma}

\begin{lemma}[\cite{Berman}]
  In the situation of Definition~\ref{DefCharges}, for $u\in A^*$ and $v\in A$ with $\chrg{u}{v}>0$, we have 
 \[w^2(u) - w^2(N(u,A)\backslash\{v\}) \geq 2\cdot\chrg{u}{v}\cdot w(v). \]\label{LemPropPositiveCharges}
\end{lemma}
\begin{lemma}[\cite{Berman}]
Let $G=(V,E)$ be $d$-claw free and $w:V\rightarrow\mathbb{R}_{>0}$. Let further $A^*$ be an independent set in $G$ of maximum weight and let $A$ be independent in $G$ with the property that no claw improves $w^2(A)$.
Then for each $v\in A$, we have \[\sum_{u\in A^*:\chrg{u}{v}>0}\chrg{u}{v}\leq \frac{w(v)}{2}. \]\label{LemBoundCharges}\end{lemma}
By combining Corollary~\ref{CorBoundwAstar} with the previous lemmata, one obtains Theorem~\ref{TheoApproxFactor}, which states an approximation guarantee of $\frac{d}{2}$.
\begin{theorem}[\cite{Berman}]
 Let $G=(V,E)$ be $d$-claw free and $w:V\rightarrow\mathbb{R}_{>0}$. Let further $A^*$ be an independent set in $G$ of maximum weight and let $A$ be independent in $G$ with the property that no claw improves $w^2(A)$.
 Then \[w(A^*)\leq \sum_{u\in A^*} \frac{w(N(u,A))}{2}+\sum_{u\in A^*: \chrg{u}{n(u)}>0} \chrg{u}{n(u)}\leq \frac{d}{2}\cdot w(A). \]\label{TheoApproxFactor}
\end{theorem}

After having recapitulated the results from \cite{Berman} that we will reemploy in our analysis, we are now prepared to study our algorithm that takes into account a broader class of local improvements. More precisely, similar as in the $\frac{k+1}{3}+\epsilon$-approximation algorithm for the unweighted $k$-Set Packing Problem by F\"urer and Yu \cite{FurerYu}, we want to consider local improvements of logarithmic size. These will correspond to cycles in some auxiliary graph. To this end, we need a statement somewhat similar to Lemma~$3.2$ from \cite{berman1994approximating} stating that if a graph possesses considerably more edges than vertices, we can find binoculars (i.e. subgraphs consisting of two cycles connected by a path or sharing a consecutive sequence of edges or a vertex) of logarithmic size. However, as we are in a weighted setting, we have to prove a slightly different statement.
\section{The existence of local improvements of logarithmic size\label{SecLogSizeImprovements}}
In this section, we aim at obtaining similar results about the existence of local improvements of logarithmic size as for unit weights. In this unweighted setting, one of the main steps is to bound the number of vertices (in the conflict graph of a $d-1$-Set Packing instance) in an optimum solution $A^*$ that have degree one or two to the current solution $A$ \cite{FurerYu}. The intuition behind that might be that if all vertices in $A^*$ had degree at least $3$ to $A$, then, as $d$-claw-freeness of $G$ implies that each vertex in $A$ can have degree at most $d-1$ to $A^*$, one would immediately get that $|A^*|\leq \frac{d-1}{3}|A|$. Now, to bound the number of vertices from $A^*$ having degree one or two to $A$, an auxiliary multi-graph is constructed, where the vertices correspond to $A$, and nodes from $A^*$ with degree one to $A$ induce a loop on the respective vertex, while nodes from $A^*$ with degree two to $A$ correspond to an edge between their neighbors \cite{FurerYu}. It is not hard to see that subgraphs of the auxiliary graph that contain more edges than vertices yield local improvements. (Note that all edges correspond to vertices from the independent set $A^*$.) This motivates the following definition:
\begin{definition}
	Let $G$ be an undirected graph, that may contain parallel edges and loops, which are counted twice towards the degree. We call a subgraph $H$ of $G$ \emph{improving} if every vertex in $H$ has degree at least $2$ and $|E(H)|>|V(H)|$ (i.e. there exists a vertex of degree at least $3$ in $H$). We call $|E(H)|$ the \emph{size} of $H$.
\end{definition}
The idea of our analysis is to do a similar construction as in the unweighted case for a subgraph in which the weights of vertices that are connected by an edge only deviate by a factor very close to $1$. In doing so, we encounter auxiliary graphs that bear a special structure. The following lemma shows that we can find improving subgraphs of logarithmic size in these.
\begin{lemma}
	Let $G$ be a finite, undirected graph on at least two vertices, let $\alpha \geq 1$ and let $w:V(G)\rightarrow\mathbb{R}_{>0}$ with the property that for each $\{u,v\}\in E(G)$, we have $\frac{w(u)}{w(v)}\in[\alpha^{-\frac{1}{2}},\alpha^\frac{1}{2}]$. Let further $t\in\mathbb{N}_{\geq 5}$ such that $\frac{2\alpha}{t-2}<1$ and let $V(G)=X\dot{\cup}Y$ such that $Y$ is stable and no vertex in $Y$ has incident loops, $((1-\frac{2\alpha}{t-2})^{-1}-1)\cdot w(Y)<w(X)$ and every $v\in X$ has degree at least $t$ in $G$. Then $G$ contains an improving subgraph of size at most $32\cdot\log(|V(G)|)$. Moreover, $G$ contains a cycle of size at most $4\cdot\log(|V(G)|)$.\label{LemFindLogarithmicImprovement}
\end{lemma}
\begin{proof}
	Note that we do not require $G$ to be simple and allow loops, which we count twice when it comes to the degree. Moreover, we regard a loop or two parallel edges as edge sets of cycles. Observe that if we have three copies of some edge, then we have found an improving subgraph and are hence done. We can, therefore, assume that this is not the case. If we only want to find a cycle, we can even assume that there are no two parallel edges and no loops in $G$. We first perform the following preprocessing steps:
	\begin{enumerate}
		\item Delete all isolated vertices from $Y$. This can only decrease the weight of $Y$ and does not harm any of the other conditions.
		\item For each vertex $y\in Y$ of degree $k\geq 2$, fix an order $x_1,\dots,x_k$ of the neighbors of $Y$ such that if some vertex $x$ occurs twice among the neighbors of $y$ (i.e. there are two parallel edges $\{x,y\}$), then the two occurrences of $x$ in the list are adjacent. Note that all neighbors of $y$ come from $X$ since $Y$ is stable and there are no loops incident to vertices in $Y$. Now, replace the edges $\{x_1,y\},\dots,\{x_k,y\}$ by the edges $\{x_1,x_2\},\{x_2,x_3\},\dots,\{x_{k-1},x_k\}$, where loops are allowed, and delete $y$ from $G$. For each of the new edges $\{x_i,x_{i+1}\}$, we have \[\frac{w(x_{i+1})}{w(x_i)}=\frac{w(x_{i+1})}{w(y)}\cdot \frac{w(y)}{w(x_i)}\leq (\alpha^\frac{1}{2})^2=\alpha\] and, analogously, $\frac{w(x_{i+1})}{w(x_i)}\geq \alpha^{-1}$.\\
		 Moreover, the degree of no vertex in $X$ can decrease by this operation: If $x$ occurs once among the neighbors of $y$, then the fact that $k\geq 2$ implies that $x$ receives at least one incident edge. If $x$ occurs twice among the neighbors of $y$, then we build a loop on $x$ counting twice towards the degree of $x$ (plus maybe some further edges). By our previous assumption, there cannot be more than $2$ parallel edges.\\
		  In case we are only looking for a cycle in $G$ of logarithmic size, then we can even assume that there are no loops and no parallel edges initially, meaning that we may introduce parallel edges, but no loops. However, if we introduce parallel edges, this means that we have found a cycle of length at most $4$ (depending on whether both of the cycle edges are newly constructed or just one), so we could even assume that there are no parallel edges when looking for a cycle in $G$.
	\end{enumerate}
	Call the set of vertices from $Y$ that survive the preprocessing $Y'$. By definition of the preprocessing and since $Y$ is stable, each vertex in $Y'$ has degree $1$.\\
	 Let $G'$ be the graph resulting from the preprocessing. For a subgraph $G''$ of $G'$ and $v\in V(G'')$, we denote the degree of $v$ in $G''$ by $\mathrm{deg}_{G''}(v)$. In particular, for $v\in W\subseteq V(G')$, $\mathrm{deg}_{G'[W]}(v)$ denotes the degree of $v$ in the subgraph of $G'$ that is induced by $W$.\\
	  Moreover, for $v\in V(G')$ and $W\subseteq V(G')\backslash\{v\}$, we use the term $\delta_{G'}(v,W)$ to refer to the set of incident edges of $v$ that have their other endpoint in $W$, and denote its cardinality by $\mathrm{deg}_{G'}(v,W):=|\delta_{G'}(v,W)|$. Note that as $v\not\in W$, $\delta_{G'}(v,W)$ does not contain any loop and $\mathrm{deg}_{G'}(v,W)$ actually is the portion of the degree of $v$ coming from edges with one endpoint in $W$.\\
	   We now have the following properties:
	\begin{equation}
	\forall \{u,v\}\in E(G'): \frac{w(u)}{w(v)}\in[\alpha^{-1},\alpha] \label{EqSimilarWeights}\end{equation}
	\begin{equation}\forall y\in Y':\mathrm{deg}_{G'}(y)=1\label{EqdegY}\end{equation}
	\begin{equation}
	w(Y')\leq w(Y)<\left(\left(1-\frac{2\alpha}{t-2}\right)^{-1}-1\right)^{-1}\cdot w(X) \label{EqWeightY}\end{equation}
	\begin{equation}
	\forall x\in X:|\mathrm{deg}_{G'}(x)|\geq t >4\label{EqDegX}
	\end{equation}
	Consider Algorithm~\ref{AlgVertexDeletion}.
	\begin{algorithm}[t]
		\DontPrintSemicolon
		$Y^0\gets Y'$\;
		\For{$i\gets 1$ \KwTo $\infty$}{
			$Y^i\gets\{x\in X\backslash\bigcup_{0\leq j\leq i-1} Y^j: \mathrm{deg}_{G'[X\backslash\bigcup_{0\leq j\leq i-1} Y^j]}(x)\leq 2\}$\;}
		\caption{Bad vertex deletion}\label{AlgVertexDeletion}
	\end{algorithm}
	Note that after at most $|V(G')|$ iterations, we have $Y^i=\emptyset$ because if $Y^i = \emptyset$, then $Y^{i+1}=Y^i=\emptyset$ by definition of the sets $Y^j$ and since all vertices in $X$ have degree $>2$ in $G'$, and it follows inductively that all subsequent sets are empty. On the other hand, each iteration where $Y^i$ is non-empty removes at least one vertex, so there can be at most $|V(G')|$ of these. Set $\bar{Y}:=\bigcup_{i=0}^\infty Y^i=\bigcup_{i=0}^{|V(G')|} Y^i$.
	 \begin{claim*}$w(\bar{Y}\backslash Y')< w(X)$.\end{claim*}Observe that this claim implies $X\backslash\bar{Y}\neq \emptyset$ because $X\backslash\bar{Y}=X\backslash(\bar{Y}\backslash Y')$ since $Y'\subseteq Y$ and $Y$ and $X$ are disjoint. Further note that as all occurring sets are finite as subsets of $V(G')$, everything is well-defined.
	\begin{claimproof}[Proof of the claim] By definition, we have $w(\bar{Y})=\sum_{i=0}^{\infty}
	w(Y^i)$, where only finitely many summands are non-zero. By \eqref{EqDegX}, the definition of $Y^i$ for $i\geq 1$ and as $V(G')\backslash X = Y' = Y^0$, we get \begin{equation}\forall i\geq 1:\forall x\in Y^i:|\delta_{G'}(x, \bigcup_{0\leq j\leq i-1} Y^j)|\geq t-2\label{EqLargeDegreeToPredecessors}.\end{equation} Moreover, by \eqref{EqdegY} and as for $i\geq 1$ and $v\in Y^i$, we have \[|\delta(v,\bigcup_{j\geq i+1} Y^j)|\leq \mathrm{deg}_{G'[X\backslash\bigcup_{0\leq j\leq i-1} Y^j]}(v)\leq 2,\] we further obtain
	\begin{equation}
	\forall i\in\mathbb{N}_0:\forall v\in Y^i:|\delta_{G'}(v,\bigcup_{j\geq i+1} Y^j)|\leq 2.\label{EqSmallDegreeToSucc}
	\end{equation}
	From this, using that in each of the following sums, only finitely many summands are non-zero, we calculate
	\begin{align*}
	w(\bar{Y})&=\sum_{i=0}^{\infty}w(Y^i) = w(Y^0)+\sum_{i=1}^{\infty}w(Y^i)= w(Y')+\sum_{i=1}^{\infty}\sum_{y\in Y^i} w(y)\\
	&\stackrel{\eqref{EqLargeDegreeToPredecessors}}{\leq} w(Y')+\frac{1}{t-2}\sum_{i=1}^{\infty}\sum_{y\in Y^i}|\delta_{G'}(y, \bigcup_{0\leq j\leq i-1} Y^j)|\cdot w(y) \\
	&= w(Y')+\frac{1}{t-2}\sum_{i=1}^{\infty}\sum_{y\in Y^i}\sum_{e=\{y,z\}\in \delta_{G'}(y, \bigcup_{0\leq j\leq i-1} Y^j)} w(y) \\
	&\stackrel{\eqref{EqSimilarWeights}}{\leq} w(Y')+\frac{1}{t-2}\sum_{i=1}^{\infty}\sum_{y\in Y^i}\sum_{e=\{y,z\}\in \delta_{G'}(y, \bigcup_{0\leq j\leq i-1} Y^j)}\alpha\cdot w(z)\\
	&= w(Y')+\frac{\alpha}{t-2}\sum_{i=1}^{\infty}\sum_{y\in Y^i}\sum_{e=\{y,z\}\in \delta_{G'}(y, \bigcup_{0\leq j\leq i-1} Y^j)} w(z) \\&= w(Y')+\frac{\alpha}{t-2}\sum_{i=0}^{\infty}\sum_{z\in Y^i}|\delta_{G'}(z,\bigcup_{j\geq i+1}Y^j)|\cdot w(z)\\&\stackrel{\eqref{EqSmallDegreeToSucc}}{\leq} w(Y')+\frac{2\alpha}{t-2}\sum_{i=0}^{\infty}\sum_{z\in Y^i}w(z)=w(Y')+\frac{2\alpha}{t-2}\cdot w(\bar{Y}).
	\end{align*} From this, we get $(1-\frac{2\alpha}{t-2})\cdot w(\bar{Y})\leq w(Y')$ and our assumption that $\frac{2\alpha}{t-2}<1$ further yields
	\[w(\bar{Y})\leq \left(1-\frac{2\alpha}{t-2}\right)^{-1}\cdot w(Y').\]
	As $Y'\subseteq \bar{Y}$, this leads to  \begin{align*}w(\bar{Y}\backslash Y')&\leq \left(\left(1-\frac{2\alpha}{t-2}\right)^{-1}-1\right)\cdot w(Y')\\ &\stackrel{\eqref{EqWeightY}}{<}\left(\left(1-\frac{2\alpha}{t-2}\right)^{-1}-1\right)\cdot\left(\left(1-\frac{2\alpha}{t-2}\right)^{-1}-1\right)^{-1}\cdot w(X)=w(X), \end{align*} i.e. $w(\bar{Y}\backslash Y')<w(X).$ This proves the claim.\end{claimproof}
	Let $X^*:=X\backslash\bar{Y}$. Assume there were a vertex $x\in X^*$ such that $\mathrm{deg}_{G'[X^*]}(x)\leq 2$. As $|V(G')|$ is finite, $x$ has only finitely many neighbors and in particular, there are only finitely many sets $Y^j$ containing a neighbor of $x$. Hence, there exists $i$ such that $\mathrm{deg}_{G'[X\backslash\bigcup_{0\leq j\leq i} Y^j]}(x)\leq 2$. Pick $i\in\mathbb{N}_0$ minimum with this property. But then, as $x\in X^*$ and, therefore, $x\in X\backslash \bigcup_{0\leq j\leq i} Y^j$, we get $x\in Y^{i+1}\subseteq \bar{Y}$, a contradiction. Consequently, every vertex in $X^*$ has degree at least $3$ in $G'[X^*]$. By our definition of the degree of loops, this implies that $|E(G'[X^*])|\geq \frac{3}{2}\cdot |X^*|$ because when summing up all degrees, each edge is counted twice. If $|X^*|=1$, then the unique vertex in $X^*$ must have at least two incident loops, and we have found cycle and an improving subgraph of size at most $2$.  Otherwise, we can apply Lemma~$3.2$ from \cite{berman1994approximating}, which tells us that any graph $G=(V,E)$ with $|E|\geq\frac{s+1}{s}\cdot |V|$ for some integer $s\geq 1$ contains an improving subgraph with less than $4\cdot s\cdot\log(|V|)$ nodes. In particular, by removing edges from this subgraph until the number of edges is by excactly one larger than the number of vertices, we find an improving subgraph of size at most $4\cdot s\cdot\log(|V|)$. Hence, we know that $G'[X^*]$ contains an improving subgraph $H'$ with at most $8\cdot\log(|X^*|)$ edges. Also, the proof of Lemma $3.1$ from \cite{berman1994approximating} (using the stronger assumption that every vertex has degree at least $3$) implies that $G'[X^*]$ contains a cycle $C'$ of length at most $2\cdot\log(|X^*|)$. Anyways, we could also just take some cycle in $H'$, resulting in a slightly worse size bound.\\
	 We want to transform $C'$ into a cycle  $C$ in $G$ and $H'$ into an improving subgraph $H$ in $G$. To this end, call an edge in $G'$ \emph{dashed} if it was introduced by the second preprocessing step, and \emph{solid} otherwise. For a dashed edge $e$, let $y_e\in Y$ be the vertex it corresponds to.
	We first consider $C'$, and then $H'$.\\
	\textbf{$C'$:} As long as there is a dashed edge left in $C'$, do the following: Let $y$ be the vertex the dashed edge corresponds to. Consider the connected components of $C'$ induced by the dashed edges corresponding to $y$, where we ignore components that are isolated vertices. Then each component is a path because all of the dashed edges corresponding to $y$ form a path (recall that we could assume that we do not create loops).\\ In case there is only one component, let $x_1$ and $x_2$ be its endpoints and let $P$ be the $x_1$-$x_2$-path in $C'$ consisting of the edges that are not dashed edges induced by $y$.\\ In case there are at least two connected components, pick $x_1$ and $x_2$ as two endpoints of two neighboring components such that $x_1$ and $x_2$ are connected by a path consisting of edges from $C'$ not induced by $y$. Define $P$ to be this path.\\ In either case, replace the arc $C'-P$ by $\{x_1,y\}$, $\{x_2,y\}$. In doing so, we replace a sub-path of $C'$ containing all dashed edges incident to $y$ by a path of length $2$ via $y$. As $y$ did not occur among the vertices of $C'$ before, this maintains the property that $C'$ is a cycle. Also, we do not introduce any new dashed edges and the only new vertex added in such a step is $y$. Hence, the procedure terminates and we add at most one new vertex per dashed edge in $C'$, meaning that we obtain a cycle $C$ in $G$ of size \begin{align*}|E(C)|&=|V(C)|\leq |V(C')|+|E(C')|=2\cdot |V(C')|\leq \max\{2,4\cdot\log(|X^*|)\}\\&\leq 4\cdot\log(|V(G)|).\end{align*}
	\\
	\textbf{$H'$:} Perform the following postprocessing steps:
	\begin{enumerate}
		\item Determine all pairs $(y,x)$ with $x\in X^*$ and $y\in Y$ such that $H'$ contains a dashed edge incident to $x$ that corresponds to $y$. For each such pair, add all copies of $\{x,y\}$ existing in $G$ to $H'$.
		\item Remove all dashed edges from $H'$.
		\item Delete all vertices from $X^*$ that have degree $1$ (in $H'$) now.
\end{enumerate} 
Call the resulting graph $H$. By definition, it is a subgraph of $G$. As we have at most two copies of each edge and add at most $1$ new vertex per dashed edge in $H'$, each dashed edge leads to at most four new edges (two per endpoint). If $m_d$ denotes the number of dashed edges in $H'$, then the number of edges of $H$ is bounded by 
\begin{align*}|E(H)|&\leq |E(H')|-m_d+4\cdot m_d\leq |E(H')|+3m_d\leq 4|E(H')|\\&\leq 4\cdot\max\{2,8\cdot\log(|X^*|)\}\leq 32\cdot\log(|V(G)|).\end{align*}
We need to see that $H$ is improving. Note that as soon as we have shown this, we also get $|V(H)|\leq |E(H)|$. First, there cannot be an isolated vertex from $X^*$ in $H$ after the deletion of dashed edges because if a vertex in $X^*$ has an incident dashed edge in $H'$, then it has an incident edge to $Y$ in $H$. Furthermore, note that each vertex $x$ deleted in the third step can only have had a neighbor in $Y$: If not, $x$ must have had at least one incident solid edge (as we only add new edges to $Y$) and hence exactly one since it would not be deleted in the third step otherwise. As $H'$ was improving, $x$ must have had degree at least $2$ (and therefore an incident dashed edge) in $H'$. But now, the fact that $x$ had at least one incident dashed edge implies that we have added at least one new edge incident to a vertex in $Y$, and together with the solid edge, this produces a degree of at least $2$, a contradiction. In particular, removing the vertices of degree $1$ from $X^*$ can only decrease the degree of vertices in $Y$. Hence, all vertices in $V(H)\cap X$ have degree at least $2$ because there is no isolated vertex after step $2$ and step $3$ deletes all vertices of degree $1$ from $X$.\\
 Next, we want to see that all vertices in $V(H)\cap Y$ have degree at least $2$ in $H$. To this end, note the degree of a vertex $x\in X$ can only drop to $1$ in the second step if all of its incident edges are dashed and correspond to the same vertex in $y$, to which it has exactly one edge. As we add at most two dashed edges (and no dashed loop) incident to $x$ and corresponding to $y$ in such a case, the incident edges of $x$ in $H'$ are precisely the two dashed edges corresponding to $y$.\\ Initially, each vertex $y\in Y\cap V(H)$ has degree at least $2$ since the dashed edge in $H'$ it corresponds to has at least two endpoints or is a loop corresponding to two parallel edges we add. We know that whenever a neighbor $x$ of $y$ is deleted, it had precisely two incident edges in $H'$ and both of them are dashed non-loop edges corresponding to $y$. Consider a maximal path in $H'$ all edges of which are dashed and correspond to $y$ and all inner vertices of which are deleted from $H'$. Consider the endpoints of the path. If one of them is deleted, both incident edges in $H'$ are dashed and correspond to $y$. Moreover, as after removing the loops, the dashed edges corresponding to $y$ do not contain a cycle by construction, we could continue our path in this case, leading to a contradiction. Hence, both endpoints of the path are not deleted and distinct by definition of a path and have incident dashed edges corresponding to $y$ in $H'$. Hence, $y$ becomes connected to both of them and has degree at least $2$ in $H$.\\
It remains to see that at least one vertex in $H$ has degree at least $3$. We have already seen that we only delete vertices from $X^*$ that had degree $2$ before. Now, suppose there is $x\in X^*$ of degree at least $3$ in $H'$ such that the degree of $x$ drops to $2$ in $H$. (We are done if the degree remains at least $3$ and we do not generate isolated vertices and the degree cannot drop to $1$ as we had degree $\geq 3$ before.) We distinguish two cases:\\
\begin{enumerate}
	\item $x$ has $2$ incident dashed edges corresponding to the same $y$. At most one of them is a loop since we assumed to have at most two copies per edge initially, so $x$ occurred at most twice in the list. In case of a loop, we get two edges between $x$ and $y$ and an additional edge from $y$ to the other endpoint of the non-loop. In case none of the two edges is a loop, its two other endpoints are distinct since the non-loop dashed edges corresponding to $y$ do not induce a cycle, so again $y$ has degree at least $3$ (1 edge to $x$ and to each of the other endpoints). By considering maximal paths all inner vertices of which only have dashed edges incident to $y$ and that are deleted starting in $x$ with the non-loop dashed edges corresponding to $y$, we see that $y$ still has degree $\geq 3$ after the deletions.
	\item No two dashed edges of $x$ correspond to the same vertex in $y$. All solid edges of $x$ remain as we do not delete endpoints of solid edges as we have seen.
	A dashed loop is replaced by two incident edges to the corresponding $y$, and a dashed edge by one incident edge to the corresponding $y$. Hence, the degree of $x$ cannot decrease.
\end{enumerate}
As a consequence, we can conclude that $H$ yields an improving subgraph of logarithmic size \mbox{$|E(H)|\leq 32\cdot\log(|V(G)|)$} as claimed.\end{proof}
 \section{Improving the Approximation Factor\label{SecAlgo}}
 \subsection{The Algorithm LogImp}
 \begin{definition}[Local improvement]
 	Given a $d$-claw free graph $G=(V,E)$, a (positive) weight function $w:V\rightarrow\mathbb{R}_{>0}$ and an independent set $A\subseteq V$, we call an independent set $X\subseteq V$ a \emph{local improvement} of $w^2(A)$ if \[w^2(A\backslash N(X,A)\cup X) > w^2(A).\]\label{DefLocalImprovement}
 \end{definition}
 \begin{proposition}
  Let $G$, $w$ and $A$ be as in Definition~\ref{DefLocalImprovement}. If $X$ is a local improvement of $w^2(A)$, then $A\backslash N(X,A)\cup X$ is independent in $G$.\label{PropMaintainIndependentSet}
 \end{proposition}
% \begin{proof}
%  As $A$ and $X$ are independent, no pair of vertices from $A\backslash N(X,A)$  or from $X$ can be adjacent. By definition of $N(X,A)$, there is no edge in $E$ connecting $A\backslash N(X,A)$ to $X$, which completes the proof.
% \end{proof}
\begin{proposition}
Let $G$, $w$ and $A$ be as in Definition~\ref{DefLocalImprovement}. Then an independent set $X$ constitutes a local improvement of $A$ if and only if we have
$w^2(N(X,A))<w^2(X)$.\label{PropEquivDefLocalImpr}
\end{proposition}
\begin{proof}
 By Definition~\ref{DefNeighborhood}, we have $N(X,A)\subseteq A$ and $(A\backslash N(X,A))\cap X=\emptyset$, so 
 \begin{align*}w^2(A\backslash N(X,A)\cup X)&=w^2(A\backslash N(X,A))+w^2(X)\\&=w^2(A)-w^2(N(X,A))+w^2(X),\end{align*} implying the claim.
\end{proof}
\begin{definition}[Claw-shaped local improvement]
	We call a local improvement $X$ \emph{claw-shaped} if $|X|=1$ and $N(X,A)=\emptyset$ or if there is $v\in A$ such that $\{v\}\cup X$ induces a $|X|$-claw in $G$ centered at $v$, that is if $X$ is the set of talons of some claw in $G$ centered in $A$, if the center is non-empty.
	\end{definition}
\begin{definition}[Circular improvement]
	Let two fixed maps \begin{itemize} \item $n:\{u\in V\backslash A: N(u,A)\neq\emptyset\}\rightarrow A$ mapping $u$ to an element of $N(u,A)$ of maximum weight and \item $n_2:\{u\in V\backslash A: |N(u,A)|\geq 2\}\rightarrow A$ mapping $u$ to an element of $N(u,A)\backslash\{n(u)\}$ of maximum weight\end{itemize} be given.\\ We call a local improvement $X\subseteq V\backslash A$ \emph{circular} if there exists\\ $U\subseteq X\cap\{u\in V\backslash A: |N(u,A)|\geq 2\}$ with $|U|\leq 4\cdot \log(|V|)$ such that 
	\begin{romanenumerate}
		\item $C:=(\bigcup_{u\in U}\{n(u), n_2(u)\}, \{e_u=\{n(u), n_2(u)\}, u\in U\})$ is a cycle, where the edge set is considered as a multiset and two parallel edges are regarded as forming a cycle of length $2$.
		\item If we define $Y_v:=\{x\in X\backslash U: n(x)=v\}$ for $v\in A$, then $|Y_v|\leq d-1$ for all $v\in A$ and moreover, \mbox{$X=U\cup\bigcup_{v\in V(C)} Y_v$}.
		\item For each $u\in U$:
		\begin{align*}w^2(u)+\frac{1}{2}\cdot w^2(Y_{n(u)}\cup Y_{n_2(u)})&> \frac{w^2(n(u))+w^2(n_2(u))}{2}\\\phantom{=}&+w^2(N(u,A)\backslash\{n(u),n_2(u)\})\\\phantom{=}&+\frac{1}{2}\cdot\sum_{x\in Y_{n(u)}} w^2(N(x,A)\backslash\{n(u)\})\\\phantom{=}&+\frac{1}{2}\cdot\sum_{x\in Y_{n_2(u)}} w^2(N(x,A)\backslash\{n_2(u)\})\end{align*}
		\end{romanenumerate}\label{DefCircular}
	\end{definition}
 The remainder of Section~\ref{SecAlgo} is now dedicated to the analysis of our algorithm LogImp (Algorithm~\ref{LocalImprovementAlgo}) for the Maximum Weight Independent Set Problem in $d$-claw free graphs for $d\rightarrow \infty$. Note that in LogImp, we can always first check for a claw-shaped improvement and if none exists, we know that $A$ is maximal and that we can define maps $n$ (with domain $V\backslash A$) and $n_2$ meeting the requirements of Definition~\ref{DefCircular}. \\
 \begin{algorithm}[t]
 		\DontPrintSemicolon
\KwIn{an undirected $d$-claw free graph $G=(V,E)$ and a (positive) weight function $w:V\rightarrow\mathbb{R}_{>0}$\;}
\KwOut{an independent set $A\subseteq V$\;}
 	$A\gets\emptyset$ \;
 	\While{there exists a claw-shaped or circular local improvement $X$ of $w^2(A)$}{
 	$A\gets A\backslash N(X,A)\cup X$	}
 	\Return A\;
 	\caption{The algorithm LogImp.}\label{LocalImprovementAlgo}
 \end{algorithm}
 The main result of this section is given by the following theorem:
 \begin{theorem}
 	For any $\delta >0$, there exists $d_\delta\in\mathbb{N}$ with the following property:
 If $A^*$ is an optimum solution to the MWIS in a $d$-claw free graph $G$ for some $d\geq d_\delta$ and $A$ denotes the solution returned by LogImp, then we have \[w(A^*)\leq \frac{d-1+\delta}{2}\cdot w(A).\] \label{TheoMainResult}
 \end{theorem}
 First, note that LogImp is correct in the sense that it returns an independent set. This follows immediately from the fact that we maintain the property that $A$ is independent throughout the algorithm, because $\emptyset$ is independent and Proposition~\ref{PropMaintainIndependentSet} tells us that none of our update steps can harm this invariant.\\
 Next, observe that LogImp is guaranteed to terminate, since no set $A$ can be attained twice, given that $w^2(A)$ strictly increases in each iteration of the while-loop, and there are only finitely many possibilities. Furthermore, each iteration runs in quasi-polynomial time (considering $d$ a constant) because there are only $\mathcal{O}(|V|^{4\cdot d\cdot\log(|V|)})$ many possible choices for $X$ and we can check in polynomial time whether a given one constitutes a local improvement, and whether it is of one of the two allowed shapes. More precisely, to check whether a local improvement $X$ is claw-shaped, we first check whether $|X|\leq d-1$, and whether $|X|=1$ or all vertices in $X$ have a common neighbor in $A$. To check whether $X$ is circular, we can try all $2^{4\cdot d\cdot\log(|V|)}=|V|^{4d}$ possible choices for $U$. For each choice of $U$, the partition into the sets $Y_v$ is uniquely determined by the map $n$ (although it might of course be impossible to find one).\\ In order to achieve a polynomial number of iterations, we scale and truncate the weight function as explained in \cite{ChandraHalldorsson} and \cite{Berman}. Given a constant $N>1$, we first  compute a greedy solution $A'$ and rescale the weight function $w$ such that $w(A')=N\cdot |V|$ holds. Recall that the greedy algorithm is known to yield a $d-1$-approximation, and note that this property is preserved when rescaling the weight function. After the rescaling, we delete vertices $v$ of truncated weight $\lfloor w(v) \rfloor = 0$ and run LogImp with the integral weight function $\lfloor w\rfloor$. In doing so, we know that $\lfloor w\rfloor^2(A)$ equals zero initially and must increase by at least one in each iteration. On the other hand, at each point, we have \[\lfloor w\rfloor ^2(A)\leq w^2(A)\leq (w(A))^2\leq (d-1)^2w^2(A')=(d-1)^2\cdot N^2\cdot |V|^2, \]which bounds the total number of iterations by the latter term. Finally, if $r>1$ specifies the approximation guarantee achieved by LogImp, $A$ denotes the solution it returns and $A^*$ is an independent set of maximum weight with respect to the original respectively the scaled, but untruncated weight function $w$, we know that \[r\cdot w(A)\geq r\cdot\lfloor w\rfloor (A)\geq\lfloor w\rfloor(A^*)\geq w(A^*)-|A^*|\geq w(A^*)-|V|\geq  \frac{N-1}{N}\cdot w(A^*), \]so the approximation ratio increases by a factor of at most $\frac{N}{N-1}$.\\
 Before we dive into the analysis of LogImp, we point out that Theorem~\ref{TheoMainResult}, combined with the previous considerations, already implies Theorem~\ref{TheoSequenceEpsilons}:
 \begin{proof}[Proof of Theorem~\ref{TheoSequenceEpsilons}, assuming Theorem~\ref{TheoMainResult}]
 Assume that Theorem~\ref{TheoMainResult} holds. For \mbox{$\delta\in(0,1]$}, define \begin{align*}d^*_\delta:=\min\{n\in\mathbb{N}_{\geq 3}: \forall d\geq n: &\text{ LogImp yields a $\frac{d-1+\delta}{2}$-approximation for the}\\&\text{ MWIS in $d$-claw free graphs}\}.\end{align*} Note that by definition, for $0<\delta_1\leq \delta_2\leq 1$, we have $d^*_{\delta_1}\geq d^*_{\delta_2}$. Moreover, as we have seen that LogImp has to terminate eventually and that when it does, there is no claw-shaped improvement of $w^2(A)$, we can apply Theorem~\ref{TheoApproxFactor} to conclude that $d^*_1 = 3$. Hence, for each $d\geq 3$, the set $\{\delta\in(0,1]: d\geq d^*_{\delta}\}$ is non-empty and we can define \[\epsilon'_d:=\inf\{\delta\in(0,1]: d\geq d^*_{\delta}\}.\] By definition of the infimum, we know that for each $d\geq 3$, LogImp yields a $\frac{d-1+\epsilon'_d}{2}$-approximation for the MWIS in $d$-claw free graphs. On the other hand, we know that the algorithm from \cite{NeuwohnerLipics} gives a $\frac{d}{2}-\frac{1}{M}$-approximation, where $M:=63,700,992$. Hence, by either running both algorithms and taking the better one of the two solutions they produce, or also checking for all local improvements of $w^2(A)$ of constant size at most $d-1+(d-1)^2$ in each iteration of LogImp, we know that for each $d\geq 3$, we have a  $\frac{d-1+\epsilon''_d}{2}$-approximation for the MWIS in $d$-claw free graph, where $\epsilon''_d:=\min\{\frac{M-1}{M},\epsilon'_d\}$. Hence, if we choose $N_d:=2\cdot(d-1)\cdot d\cdot M+1$ and apply LogImp and the algorithm from \cite{NeuwohnerLipics}, or the extended version of LogImp, after performing the respective weight scaling as explained before, we obtain an approximation guarantee of 
 \begin{align*}\frac{N_d}{N_d-1}\cdot \frac{d-1+\epsilon''_d}{2}&=\left(1+\frac{1}{N_d-1}\right)\cdot  \frac{d-1+\epsilon''_d}{2}\\&= \left(1+\frac{1}{2\cdot(d-1)\cdot d\cdot M}\right)\cdot\frac{d-1+\epsilon''_d}{2}\\&=\dfrac{d-1+\frac{1}{2\cdot d\cdot M}+\left(1+\frac{1}{2\cdot(d-1)\cdot d\cdot M}\right)\cdot \epsilon''_d}{2}.\end{align*}
  As a consequence, we can choose \begin{align*}\epsilon_d &:= \frac{1}{2\cdot d\cdot M}+ \left(1+\frac{1}{2\cdot(d-1)\cdot d\cdot M}\right)\cdot\epsilon''_d\\&\leq \frac{1}{2\cdot d\cdot M}+\left(1+\frac{1}{2\cdot(d-1)\cdot d\cdot M}\right)\cdot\frac{M-1}{M}\\&\leq \frac{1}{2\cdot d\cdot M}+\frac{M-1}{M}+\frac{1}{2\cdot (d-1)\cdot d\cdot M} <1\end{align*} and get a quasi-polynomial time $\frac{d-1+\epsilon_d}{2}$-approximation for each $d\geq 3$. (Moreover, by the way we ensure a polynomial running time for weighted $k$-Set Packing in Section~\ref{SecPolyTime}, it is clear that the same arguments would also apply to the modified version of LogImp.) By Theorem~\ref{TheoMainResult}, we know that for each $\delta\in(0,1)$,  $d^*_{\frac{\delta}{3}}$ is finite, and hence, for $d\geq\max\{d^*_{\frac{\delta}{3}},\lceil \frac{1}{\delta} \rceil\}$, we know that $\epsilon''_d\leq \epsilon'_d\leq \frac{\delta}{3}$. This leads to \[\epsilon_d= \frac{1}{2\cdot d\cdot M}+ \left(1+\frac{1}{2\cdot(d-1)\cdot d\cdot M}\right)\cdot\epsilon''_d\leq \frac{\delta}{2M}+2\cdot \epsilon''_d\leq \frac{\delta}{3}+2\cdot\frac{\delta}{3}= \delta,\] proving that $\lim_{d\rightarrow\infty}\epsilon_d = 0$.
 \end{proof}
As a consequence, we can focus on proving Theorem~\ref{TheoMainResult} for the main analysis.
 \subsection{Analysis of the Performance Ratio}
 We now move to the analysis of the approximation guarantee. Denote some optimum solution by $A^*$ and denote the solution found by LogImp by $A$. Observe that by positivity of the weight function, $A$ must be a maximal independent set, as adding a vertex would certainly yield a claw-shaped local improvement of $w^2(A)$. Fix maps $n$ and $n_2$ as in Definition~\ref{DefCircular} (and observe that the domain of $n$ is $V\backslash A$).\\
Our goal is to prove Theorem~\ref{TheoMainResult}, i.e.\ to show that for any $\delta >0$, there exists $d_\delta\in \mathbb{N}$ such that for $d\geq d_\delta$, LogImp produces a $\frac{d-1+\delta}{2}$-approximation.
 We use some notation as well as most of the analysis of the algorithm SquareImp by Berman.
 In particular, we employ the same definition of neighborhoods and charges. Observe that this is well-defined as we have seen that the solution $A$ returned by our algorithm must constitute a maximal independent set in the given graph.\\
 Fix $0<\delta<1$. (Note that the statement for $\delta\geq 1$ follows from the statement for any smaller $\delta$, or from Theorem~\ref{TheoApproxFactor}.)  For our analysis, we need two auxiliary parameters, $\epsilon'$ and $\tilde{\epsilon}$. We choose them to be $\tilde{\epsilon}:=\frac{\delta}{2}$ and $\epsilon':=\frac{\delta^2}{2500}$. Additionally, let $d_\delta:= \frac{200,000}{\delta^3}+1$. These choices satisfy a bunch of conditions that pop up during our analysis at some point and are listed (and proven) in Appendix~\ref{AppendixInequalities}. We remark that since we are more interested in the qualitative statement anyways, our constants are not optimized, but rather chosen in a way that simplifies the proof of the conditions we want them to fulfill.
 %TODO: check that all of these constraints are actually used, add proof in appendix
 
The first step of our analysis is to examine the structure of the neighborhoods $N(v,A^*)$ of vertices $v\in A$. To this end, the following definition is required:
\begin{definition}[$T_v$ \cite{NeuwohnerLipics}]
 For $v\in A$, we define $T_v:=\{u\in A^*: \chrg{u}{v}>0\}.$
\end{definition}

\begin{definition}[Contribution \cite{NeuwohnerLipics}]
 Define a contribution map \\$\mathrm{contr}:A^*\times A\rightarrow\mathbb{R}_{\geq 0}$ by setting
 \[\mathrm{contr}(u,v):=\begin{cases}\max\left\{0,\frac{w^2(u)-w^2(N(u,A)\backslash\{v\})}{w(v)}\right\} &, \text{ if $v\in N(u,A)$}\\
                         0 &,\text{ else.}
                        \end{cases}\]
\end{definition}
\begin{proposition}[\cite{NeuwohnerLipics}]
 For each $v\in A$, we have $\sum_{u\in A^*}\mathrm{contr}(u,v)\leq w(v)$.\label{PropUpperBoundContr}
\end{proposition}
\begin{proof}
	If $v\in A^*$, this is true, because we get $N(v,A^*)=N(v,A)=\{v\}$ and $\mathrm{contr}(v,v)=w(v)$ in this case.\\
	If $v\not\in A^*$, the set $T$ of vertices sending positive contributions to $v$ constitutes the set of talons of a claw centered at $v$ and $\sum_{u\in T}\mathrm{contr}(u,v)>w(v)$ would imply that $T$ constitutes a local improvement of $w^2$.
\end{proof}
\begin{proposition}[\cite{NeuwohnerLipics}]
 For each $u\in A^*$, we have \[\sum_{v\in A}\contr{u}{v}\geq \contr{u}{n(u)}\geq 2\cdot\chrg{u}{n(u)}.\]\label{PropContrCharge}
\end{proposition}
\begin{proof}
	The first inequality follows by non-negativity of the contribution, which also implies the second inequality in case $\chrg{u}{n(u)}\leq 0$. If $\chrg{u}{n(u)}>0$, Lemma~\ref{LemPropPositiveCharges} provides the desired statement.
\end{proof}
Following \cite{NeuwohnerLipics}, we begin by classifying the vertices in $T_v$. Recall that we have fixed maps $n$ and $n_2$ according to Definition~\ref{DefCircular} for the analysis.
\begin{definition}[Single vertex]
	Let $0<\beta<1$. For $v\in A$, we call a vertex $u\in T_v$ \emph{$\beta$-single} if 
	\begin{romanenumerate}
		\item $\frac{w(u)}{w(v)} \in [1-\beta, 1+\beta]$ and 
		\item $w(N(u,A))\leq (1+\beta)\cdot w(v)$.
	\end{romanenumerate}
	
\end{definition}
\begin{definition}[Double vertex]
	Let $0<\beta<1$.	For $v\in A$, we call a vertex $u\in T_v$ \emph{$\beta$-double} if $|N(u,A)|\geq 2$ and for $v_1:=v$ and $v_2:=n_2(u)$, the following properties hold:
	\begin{romanenumerate}
		\item $\frac{w(u)}{w(v_1)}\in[1-\beta,1+\beta]$
		\item $\frac{w(v_2)}{w(v_1)}\in [1-\beta, 1]$ and
		\item $(2-\beta)\cdot w(v_1)\leq w(N(u,A))<2\cdot w(u)$.
	\end{romanenumerate}
	
\end{definition}
The following lemma generalizes Lemma $18$ from \cite{NeuwohnerLipics}.
\begin{lemma}
	Let $v\in A$, let $u\in T_v$ and let $0<\gamma<1$. Then we have one of the following:
	\begin{romanenumerate}
		\item $u$ is $\sqrt{\gamma}$-single.
		\item $u$ is $\sqrt{\gamma}$-double.
		\item $\contr{u}{v}-2\cdot\chrg{u}{v}\geq \gamma\cdot \max\left\{w(v),\frac{w(u)}{2}\right\}$. (In particular, this means that $u$ is $\frac{\gamma}{2}$-contributive, a term that is introduced in Definition~\ref{DefContributive}.)
	\end{romanenumerate}
	\label{LemClassificationTv}
\end{lemma}
\begin{proof}
	Pick $u\in T_v$. First, assume that $\contr{u}{v}-2\cdot\chrg{u}{v}\geq \gamma\cdot w(v)$. If additionally $w(v)\geq \frac{w(u)}{2}$, we are done, so assume this is not the case. As by maximality of $w(v)$ in $N(u,A)$, we have \begin{align*}(\contr{u}{v}-2\cdot\chrg{u}{v})\cdot w(v) &\geq w^2(u)-w^2(N(u,A)\backslash\{v\})\\&\phantom{=}-(2\cdot w(u)-w(N(u,A)))\cdot w(v)\\
	&\geq w^2(u)-w(v)\cdot (w(N(u,A))-w(v))\\&\phantom{=}-(2\cdot w(u)-w(N(u,A)))\cdot w(v)\\
	&= (w(u)-w(v))^2\geq \frac{w(u)}{2}\cdot w(v)
	\end{align*} since $w(v) < \frac{w(u)}{2}$, we are again done. Therefore, assume that \begin{equation}
	w^2(u)-w^2(N(u,A)\backslash\{v\})-2\cdot\chrg{u}{v}\cdot w(v) < \gamma\cdot w^2(v).\label{EqNotContr}
	\end{equation} By definition of charges, we know that $v=n(u)$ is a neighbor of $u$ in $A$ of maximum weight, implying
	\begin{align}
	w^2(N(u,A)\backslash\{v\})&=\sum_{x\in N(u,A)\backslash\{v\}} w^2(x)\notag\\
	&\leq \sum_{x\in N(u,A)\backslash\{v\}}w(x)\cdot\max\{0,\max_{y\in N(u,A)\backslash\{v\}}w(y)\}\notag\\
	&=(w(N(u,A))-w(v))\cdot\max\{0,\max_{y\in N(u,A)\backslash\{v\}}w(y)\},\label{EqBoundsWsqNeighborhood1}
	\end{align}
	where $\max\emptyset:=-\infty$. By \eqref{EqNotContr}, we obtain
	
	\begin{alignat}{3}
	& \quad w^2(u)-w^2(N(u,A)\backslash\{v\})\quad&< &\quad 2\cdot\chrg{u}{v}\cdot w(v)+\gamma\cdot w^2(v)\notag\\
	\Leftrightarrow &\quad w^2(u)-w^2(N(u,A)\backslash\{v\}) \quad&<&\quad (2\cdot w(u)-w(N(u,A)))\cdot w(v)\notag\\
	&&&\quad+\gamma\cdot w^2(v)\notag\\
	\Leftrightarrow &\quad w^2(u)+w^2(v)-w^2(N(u,A)\backslash\{v\}) \quad&<&\quad (2\cdot w(u)+w(v)-w(N(u,A)))\cdot w(v)\notag\\
	&&&\quad+\gamma\cdot w^2(v)\notag,\end{alignat}
	which results in 
	\[ (w(u)-w(v))^2-w^2(N(u,A)\backslash\{v\})+(w(N(u,A))-w(v))\cdot w(v) < \gamma\cdot w^2(v). \]

	Applying \eqref{EqBoundsWsqNeighborhood1} yields
	\begin{equation}(w(u)-w(v))^2+(w(N(u,A))-w(v))\cdot(w(v)-\max\{0,\max_{y\in N(u,A)\backslash\{v\}}w(y)\}) < \gamma\cdot w^2(v).\label{EqTogetherAtMostEpsUTimesWsqV1}\end{equation}
	As both summands in \eqref{EqTogetherAtMostEpsUTimesWsqV1} are non-negative, $v\in N(u,A)$ is of maximum weight and $w>0$, \eqref{EqTogetherAtMostEpsUTimesWsqV1} in particular implies that both \begin{align}
	\gamma\cdot w^2(v)&\geq (w(u)-w(v))^2 \label{EqUandVclose1}\text{ and}\\\gamma\cdot w^2(v)&\geq (w(N(u,A))-w(v))\cdot(w(v)-\max\{0,\max_{y\in N(u,A)\backslash\{v\}}w(y)\}).\label{EqOnceorTwiceTheWeight1}\end{align}
	From \eqref{EqUandVclose1}, we can infer that $|w(u)-w(v)|\leq\sqrt{\gamma}\cdot w(v)$, which in turn implies that  \begin{equation}\frac{w(u)}{w(v)}\in\left[1-\sqrt{\gamma},1+\sqrt{\gamma}\right].\label{EqIntervalRatioUV1}\end{equation}
	In addition to that, \eqref{EqOnceorTwiceTheWeight1} tells us that at least one of the two inequalities \begin{align} \sqrt{\gamma}\cdot w(v)&\geq w(v)-\max\{0,\max_{y\in N(u,A)\backslash\{v\}}w(y)\}\label{EqVandV2close1} \text{ or}\\ \sqrt{\gamma}\cdot w(v)&\geq w(N(u,A))-w(v)\label{EqVMakesUpAlmostWholeNeighborhood1}\end{align} must hold. If \eqref{EqVandV2close1} applies, the fact that $\gamma<1$, together with $w(v)>0$, implies that $N(u,A)\backslash\{v\}\neq \emptyset$, so let $v_2:=n_2(u)$. Then \begin{align}w(v)-w(v_2)&\leq \sqrt{\gamma}\cdot w(v)\text{ and, hence,}\notag\\ (1-\sqrt{\gamma})\cdot w(v)&\leq w(v_2)\leq w(v) \label{EqScndCondDoubleFullfilled1}\end{align} by maximality of $w(v)$ in $N(u,A)$. From this, we also get \[(2-\sqrt{\gamma})\cdot w(v)\leq w(v)+w(v_2)\leq w(N(u,A))<2\cdot w(u), \]where the last inequality follows from the fact that $u$ sends positive charges to $v$. Hence, together with \eqref{EqIntervalRatioUV1} and \eqref{EqScndCondDoubleFullfilled1}, all conditions for $u$ being $\sqrt{\gamma}$-double are fulfilled. In case \eqref{EqVMakesUpAlmostWholeNeighborhood1} holds true, we get $w(N(u,A))\leq (1+\sqrt{\gamma})\cdot w(v),$ leaving us with a vertex that is $\sqrt{\gamma}$-single by \eqref{EqIntervalRatioUV1}.
\end{proof}
  The next step is to classify the vertices in $N(v,A^*)\backslash T_v$ for $v\in A$:
  \begin{definition}[Payback vertex]
  Let $\beta > 0$. We say that $u\in A^*$ is a $\beta$-payback vertex if $w(N(u,A))\geq (2+\beta)\cdot w(u)$.
  \end{definition}
% \begin{lemma}
%  Let $\beta>0$, let $u$ be a $\beta$-payback vertex and let $v\in N(u,A)$. Then $w(N(u,A))\geq 2\cdot w(u)+\frac{\beta}{\beta+2}\cdot w(v)$.
% \end{lemma}
% \begin{proof}
%  By positivity of weights, let $\alpha:=\frac{w(u)}{w(v)}$. Then $w(N(u,A))\geq 2\cdot w(u)+\alpha\beta\cdot w(v)$. If $\alpha\geq \frac{1}{\beta +2}$, the claim follows.  On the other hand, if $\alpha<\frac{1}{\beta+2}$, by our choice of $v$, we have \[w(N(u,A))\geq w(v) = 2\cdot w(u)+w(v0-2\cdot w(u)=2\cdot w(u)+(1-2\alpha)\cdot w(v)\] and $1-2\alpha > \frac{\beta}{\beta+2}$, which again implies the claim.
% \end{proof}
\begin{definition}[Good vertex]
Let $1>\beta > 0$. We call a vertex $u\in A^*$ $\beta$-\emph{good} if $|N(u,A)|\geq 2$ and $2\cdot w(u)\leq w(N(u,A))\leq (2+\beta)\cdot w(u)$ and for $v_1:=n(u)$ and $v_2:=n_2(u)$, we have
\begin{romanenumerate}\item $w(v_2)\geq (1-\beta)\cdot w(v_1)$ and \item $\frac{w(u)}{w(v_1)}\in\left[1-\beta, \frac{1}{1-\beta}\right].$\end{romanenumerate}
\end{definition}
\begin{definition}[Contributive vertex]
Let $\beta > 0$.  We call a vertex $u\in A^*$ $\beta$-\emph{contributive} if \[\contr{u}{n(u)}\geq \beta\cdot w(u)+2\cdot\max\{0,\chrg{u}{n(u)}\}.\]\label{DefContributive}
\end{definition}

\begin{lemma}
Let $0<\gamma\leq \sqrt{2}-1$ such that $2+\gamma-\frac{1}{1-\sqrt{2\gamma}}\geq\sqrt{2\gamma}$. Let $u\in A^*$ such that $w(N(u,A))\geq 2\cdot w(u)$. Then at least one of the following applies:
 \begin{romanenumerate}
  \item $u$ is $\gamma$-payback.
  \item $u$ is $\sqrt{2\gamma}$-good.
  \item $u$ is $\gamma$-contributive.
 \end{romanenumerate}
\label{LemClassificationNoCharges}
\end{lemma}
\begin{proof}
 If $u$ is $\gamma$-payback, we are done, so assume that $w(N(u,A))< (2+\gamma)\cdot w(u)$. Let $v:=n(u)$. As $w(N(u,A))\geq 2\cdot w(u)$, we have $\chrg{u}{v}\leq 0$. If $N(u,A)=\{v\}$, then
 \[w^2(u)-w^2(N(u,A)\backslash\{v\})= w^2(u)\geq w(u)\cdot\frac{w(N(u,A))}{2+\gamma}\geq \gamma\cdot w(u)\cdot w(v)\] since $N(u,A)=\{v\}$ and $\gamma\leq \frac{1}{2+\gamma}$ because $0<\gamma\leq\sqrt{2}-1$. So $u$ is $\gamma$-contributive in this case and we can assume $|N(u,A)|\geq 2$ in the following and pick $v_2:=n_2(u)$. Then
 \begin{align*}
  w^2(u)-w^2(N(u,A)\backslash\{v\})\geq&\quad w^2(u)-w(v_2)\cdot(w(N(u,A))-w(v))\\
  \geq&\quad w^2(u)-w(v_2)\cdot ((2+\gamma)\cdot w(u)-w(v))\\
  =&\quad w^2(u)-w(v)\cdot ((2+\gamma)\cdot w(u)-w(v))\\&+(w(v)-w(v_2))\cdot ((2+\gamma)\cdot w(u)-w(v))\\
  =&\quad (w(u)-w(v))^2-\gamma\cdot w(u)\cdot w(v)\\&+(w(v)-w(v_2))\cdot ((2+\gamma)\cdot w(u)-w(v))
 \end{align*}
The first and the third summand are non-negative since real squares are non-negative and because $w(v_2)\leq w(v)$ by maximality of $w(n(u))=w(v)$ in $N(u,A)$, and $w(v)\leq w(N(u,A))< (2+\gamma)\cdot w(u)$ by our assumption that $u$ is not $\gamma$-payback.\\
 If we have $\min\{w(u),w(v)\}\leq (1-\sqrt{2\gamma})\cdot\max\{w(u),w(v)\},$ then we get \[(w(u)-w(v))^2\geq 2\gamma\cdot\max\{w(u),w(v)\}^2\geq 2 \gamma\cdot w(u)\cdot w(v),\]
 implying $w^2(u)-w^2(N(u,A)\backslash\{v\})\geq \gamma\cdot w(u)\cdot w(v)$ and hence $\gamma$-contributivity of $u$ since $u$ sends no positive charges.\\
  So assume that $\min\{w(u),w(v)\}> (1-\sqrt{2\gamma})\cdot\max\{w(u),w(v)\}$.\\ If further $w(v_2)\leq (1-\sqrt{2\gamma})\cdot w(v)$, then we obtain
 \begin{align*}
  (w(v)-w(v_2))\cdot ((2+\gamma)\cdot w(u)-w(v))&\geq \sqrt{2\gamma}\cdot w(v)\cdot \left(2+\gamma-\frac{1}{1-\sqrt{2\gamma}}\right)\cdot w(u)\\&\geq 2 \gamma\cdot w(u)\cdot w(v)
 \end{align*}
by our assumption on $\gamma$. Hence, $u$ is $\gamma$-contributive in this case. On the other hand, if $w(v_2)\geq (1-\sqrt{2\gamma})\cdot w(v)$, then all conditions for $u$ being $\sqrt{2\gamma}$-good are fulfilled.
\end{proof}

\begin{definition}[Missing neighbor]
For $v\in A$ with $|N(v,A^*)|<d-1$, we say that $v$ has $d-1-|N(v,A^*)|$ \emph{missing neighbors}.	
\end{definition}
\begin{definition}[Helpful vertex]
	We say that a vertex $u\in N(v,A^*)$ is \emph{helpful for $v$} if one of the following holds:
	\begin{itemize}
		\item $u$ is $\sqrt{\epsilon'}$-double or $\sqrt{2\epsilon'}$-good and $v\in\{n(u),n_2(u)\}$.
		\item $u$ is $\sqrt{\epsilon'}$-single and $v=n(u)$.
	\end{itemize}
	\end{definition}
We want to show that a large fraction (in terms of weight) of all vertices in $A$ has only very few neighbors that are missing, $\epsilon'$-payback, $\frac{\epsilon'}{2}$-contributive or not helpful for $v$. To this end, we need to see how a vertex $v$ \emph{profits} from neighbors of the aforementioned types.
\begin{definition}
	Define a profit map $\mathrm{profit}:A^*\times A\rightarrow \mathbb{R}_{\geq 0}$ as follows:
	\begin{enumerate}
		\item If $u\in N(v,A^*)$ is $\epsilon'$-payback, then $\mathrm{profit}(u,v):=\left(\frac{1}{2}-\frac{w(u)}{w(N(u,A))}\right)\cdot w(v)$.
		\item If $u\in N(v,A^*)$ is $\frac{\epsilon'}{2}$-contributive, but not $\epsilon'$-payback, then \[\mathrm{profit}(u,v):=\frac{1}{2}\cdot\left(\contr{u}{n(u)}-2\cdot\max\{0,\chrg{u}{n(u)}\}\right)\cdot\frac{w(v)}{w(N(u,A))}.\]
		\item If $u\in N(v,A^*)$ is $\sqrt{2\epsilon'}$-good, but not helpful for $v$, and neither $\frac{\epsilon'}{2}$-contributive nor $\epsilon'$-payback and for $v_1:=n(u)$ and $v_2:=n_2(u)$, we have \[w(v_1)+w(v_2)+(1-\epsilon')\cdot w(N(u,A)\backslash\{v_1,v_2\})\geq 2\cdot w(u),\] then \[\mathrm{profit}(u,v):=\dfrac{\frac{w(N(u,A))}{2}-w(u)}{w(N(u,A)\backslash\{v_1,v_2\})}\cdot w(v).\]Note that as $v\in N(u,A)\backslash\{v_1,v_2\}$ and all weights are positive, we do not divide by zero here.
		\item If $u\in N(v,A^*)$ is $\sqrt{2\epsilon'}$-good or $\sqrt{\epsilon'}$-double, but not helpful for $v$, and neither $\epsilon'$-payback nor $\frac{\epsilon'}{2}$-contributive and for $v_1:=n(u)$ and $v_2:=n_2(u)$, we have \[w(v_1)+w(v_2)+(1-\epsilon')\cdot w(N(u,A)\backslash\{v_1,v_2\})< 2\cdot w(u)\] (note that this condition is automatically fulfilled for $\sqrt{\epsilon'}$-double vertices since they send charges), then \[\mathrm{profit}(u,v):=\frac{1}{2}\cdot\left(\contr{u}{n(u)}-2\cdot\max\{0,\chrg{u}{n(u)}\}\right)\cdot\frac{w(v)}{w(N(u,A)\backslash\{v_1,v_2\})}.\]Again, note that as $v\in N(u,A)\backslash\{v_1,v_2\}$ and all weights are positive, we do not divide by zero here.
		\item If $u\in N(v,A^*)$ is $\sqrt{\epsilon'}$-single, but not helpful for $u$, and none of the previous cases applies to $u$, then define \[\mathrm{profit}(u,v):=\frac{1}{2}\cdot\left(\contr{u}{n(u)}-2\cdot\max\{0,\chrg{u}{n(u)}\}\right)\cdot\frac{w(v)}{w(N(u,A)\backslash\{v_1\})},\] where $v_1:=n(u)$. Again, the existence of $v\in N(u,A)\backslash\{v_1\}$ and the positivity of weights imply that we do not divide by $0$.
		\item If none of the previous cases applies, then $\mathrm{profit}(u,v):=0$.
	\end{enumerate}\label{DefProfit}
	\end{definition}
\begin{definition}
	We say that $u\in N(v,A^*)$ is \emph{profitable} for $v\in A$ if $u$ is
	\begin{itemize}
		\item $\epsilon'$-payback,
		\item $\frac{\epsilon'}{2}$-contributive,
		\item $\sqrt{2\epsilon'}$-good, but not helpful for $v$,
		\item $\sqrt{\epsilon'}$-double but not helpful for $v$, or
		\item $\sqrt{\epsilon'}$-single, but not helpful for $v$.
	\end{itemize}\label{DefProfitable}
	\end{definition}
\begin{lemma}
	Let $v\in A$ and $u\in N(v,A^*)$ such that $u$ is profitable for $v$. Then $\mathrm{profit}(u,v)\geq \frac{\epsilon'}{10}\cdot w(v)$.\label{LemProfitableVertex}
	\end{lemma}
\begin{proof}
	We distinguish the same cases as in Definition~\ref{DefProfit}.
	\begin{enumerate}
		\item As $u$ is $\epsilon'$-payback, we get $w(N(u,A))\geq (2+\epsilon')\cdot w(u)$. Hence, \begin{align*}\mathrm{profit}(u,v)&=\left(\frac{1}{2}-\frac{w(u)}{w(N(u,A))}\right)\cdot w(v)\geq \left(\frac{1}{2}-\frac{1}{2+\epsilon'}\right)\cdot w(v)\\&=\frac{\epsilon'}{2\cdot(2+\epsilon')}\cdot w(v)\geq \frac{\epsilon'}{10}\cdot w(v)\end{align*} by \eqref{const1}.
		\item As $u$ is $\frac{\epsilon'}{2}$-contributive, but not $\epsilon'$-payback, we get \[\contr{u}{n(u)}\geq 2\cdot\max\{0,\chrg{u}{n(u)}\} + \frac{\epsilon'}{2}\cdot w(u)\] and $w(N(u,A))<(2+\epsilon')\cdot w(u)$. This yields \begin{align*}\mathrm{profit}(u,v)&=\frac{1}{2}\cdot\left(\contr{u}{n(u)}-2\cdot\max\{0,\chrg{u}{n(u)}\}\right)\cdot\frac{w(v)}{w(N(u,A))}\\&\geq \frac{\epsilon'\cdot w(u)}{4}\cdot\frac{w(v)}{w(N(u,A))}\geq w(v)\cdot\frac{\epsilon'}{4}\cdot\frac{w(u)}{w(N(u,A))}\\&\geq w(v)\cdot \frac{\epsilon'}{4\cdot(2+\epsilon')}\geq \frac{\epsilon'}{10}\cdot w(v)\end{align*} by \eqref{const1}.
		\item We have $w(N(u,A))\geq 2\cdot w(u)+\epsilon'\cdot w(N(u,A)\backslash\{v_1,v_2\})$ and, therefore,
		\[\mathrm{profit}(u,v)=\dfrac{\frac{w(N(u,A))}{2}-w(u)}{w(N(u,A)\backslash\{v_1,v_2\})}\cdot w(v)\geq \frac{\epsilon'}{2}\cdot w(v).\]
		\item  First, consider the case where $u$ is $\sqrt{\epsilon'}$-double and $v\in N(u,A)\backslash\{v_1,v_2\}$, where $v_1=n(u)$ and $v_2:=n_2(u)$. Note that by Lemma~\ref{LemPropPositiveCharges},
		\[\contr{u}{v_1}\cdot w(v_1)=w^2(u)-w^2(N(u,A)\backslash\{v_1\}) >0.\] We get
		\begin{align*}
		&\phantom{=}(\contr{u}{v_1}-2\cdot\chrg{u}{v_1})\cdot w(v_1)\\&= w^2(u)-w^2(v_2)-w^2(N(u,A)\backslash\{v_1,v_2\})-w(v_1)\cdot 2\cdot\chrg{u}{v_1}\\
		&\geq w^2(u)-w^2(v_2)-\min\{w(v_2),w(N(u,A)\backslash\{v_1,v_2\})\}\cdot w(N(u,A)\backslash\{v_1,v_2\})\\&\phantom{=}-w(v_1)\cdot 2\cdot\chrg{u}{v_1}\\\
		&=w^2(u)-w^2(v_2)-\min\{w(v_2),w(N(u,A)\backslash\{v_1,v_2\})\}\cdot w(N(u,A)\backslash\{v_1,v_2\})\\&\phantom{=}-w(v_1)\cdot(2\cdot w(u)-w(N(u,A)))\\
		&=w^2(u)-w^2(v_2)-\min\{w(v_2),w(N(u,A)\backslash\{v_1,v_2\})\}\cdot w(N(u,A)\backslash\{v_1,v_2\})\\&\phantom{=}-w(v_1)\cdot(2\cdot w(u)-w(v_1)-w(v_2)-w(N(u,A)\backslash\{v_1,v_2\}))\\
		&= (w(u)-w(v_1))^2 + (w(v_1)-w(v_2))\cdot w(v_2)\\&\phantom{=}+(w(v_1)-\min\{w(v_2),w(N(u,A)\backslash\{v_1,v_2\})\})\cdot w(N(u,A)\backslash\{v_1,v_2\}).
		\end{align*}
		As $w(v_2)\leq w(v_1)$ and all weights are positive, all summands are non-negative. In case $w(v_1) \leq \frac{6}{7}\cdot w(u)$, we get
		\begin{align*}(\contr{u}{v_1}-2\cdot\chrg{u}{v_1})\cdot w(v_1)&\geq(w(u)-w(v_1))^2 \geq \frac{1}{49}\cdot w^2(u)\\&\geq \frac{1}{98}\cdot w(v_1)\cdot w(N(u,A)\backslash\{v_1,v_2\})\end{align*} since $w(N(u,A)\backslash\{v_1,v_2\})\leq w(N(u,A))<2\cdot w(u)$ as $u$ is $\sqrt{\epsilon'}$-double. In case \mbox{$w(v_1)> \frac{6}{7}\cdot w(u)$}, we must have  \[\min\{w(v_2),w(N(u,A)\backslash\{v_1,v_2\})\}\leq \frac{2}{3}\cdot w(v_1),\] because otherwise, we would get\[w(N(u,A))=w(v_1)+w(v_2)+w(N(u,A)\backslash\{v_1,v_2\})> \left(\frac{6}{7}+\frac{4}{7}+\frac{4}{7}\right)\cdot w(u)=2\cdot w(u),\] a contradiction to $u$ being $\sqrt{\epsilon'}$-double. Hence, we obtain
		\begin{align*}&\phantom{=}(\contr{u}{v_1}-2\cdot\chrg{u}{v_1})\cdot w(v_1)\\&\geq (w(v_1)-\min\{w(v_2),w(N(u,A)\backslash\{v_1,v_2\})\})\cdot w(N(u,A)\backslash\{v_1,v_2\})\\
		&\geq \frac{1}{3}\cdot w(v_1)\cdot w(N(u,A)\backslash\{v_1,v_2\}).\end{align*} 
		Therefore, in either case, given that double vertices send positive charges, we obtain \[\mathrm{profit}(u,v)\geq \frac{1}{196}\cdot w(v).\] By \eqref{const1}, this is at least $\frac{\epsilon'}{10}\cdot w(v)$.\\
		Now, consider the case where $u$ is $\sqrt{2\epsilon'}$-good. By definition, $u$ does not send positive charges. We further have 
		\[2\cdot w(v_2)\leq w(v_1)+w(v_2)< 2\cdot w(u)-(1-\epsilon')\cdot w(N(u,A)\backslash\{v_1,v_2\})\] and, therefore, 
		\[0<w(v_2)< w(u)-\frac{1-\epsilon'}{2}\cdot w(N(u,A)\backslash\{v_1,v_2\}).\] Hence,
		\begin{align*}
		\contr{u}{v_1}\cdot w(v_1)&= \max\{0,w^2(u)-w^2(v_2)-w^2(N(u,A)\backslash\{v_1,v_2\})\}
		\\&\geq w^2(u)-\left(w(u)-\frac{1-\epsilon'}{2}\cdot w(N(u,A)\backslash\{v_1,v_2\})\right)^2\\
		&\phantom{=}-(w(N(u,A)\backslash\{v_1,v_2\}))^2\\
		&\geq w(u)\cdot(1-\epsilon')\cdot w(N(u,A)\backslash\{v_1,v_2\})\\
		&\phantom{=}-\frac{5}{4}\cdot (w(N(u,A)\backslash\{v_1,v_2\}))^2.
		\end{align*}
		As $u$ is not $\epsilon'$-payback, but $\sqrt{2\epsilon'}$-good, we get \begin{align*}w(N(u,A)\backslash\{v_1,v_2\})&=w(N(u,A))-w(v_1)-w(v_2)\\&\leq w(u)\cdot (2+\epsilon'-(1-\sqrt{2\epsilon'})-(1-\sqrt{2\epsilon'})^2)\\&\leq 3\cdot\sqrt{2\epsilon'}\cdot w(u).\end{align*} 
		This implies
		\begin{align*}\contr{u}{v_1}\cdot w(v_1)&\geq w(u)\cdot\left(1-\epsilon'-\frac{15}{4}\sqrt{2\epsilon'}\right)\cdot w(N(u,A)\backslash\{v_1,v_2\})\\&\geq w(v_1)\cdot\left(1-\epsilon'-\frac{15}{4}\sqrt{2\epsilon'}\right)\cdot(1-\sqrt{2\epsilon'})\cdot w(N(u,A)\backslash\{v_1,v_2\}) .\end{align*}
		Therefore, \eqref{const2} allows us to conclude that \[\mathrm{profit}(u,v)\geq \frac{1}{2}\cdot \left(1-\epsilon'-\frac{15}{4}\sqrt{2\epsilon'}\right)\cdot(1-\sqrt{2\epsilon'})\cdot w(v)\geq \frac{\epsilon'}{10}\cdot w(v).\] 
		\item We get $\chrg{u}{n(u)}>0$ and $w(N(u,A))\leq (1+\sqrt{\epsilon'})\cdot w(n(u))$ by definition of $\sqrt{\epsilon'}$-single. For $v_1:=n(u)$, Lemma~\ref{LemPropPositiveCharges} tells us that \begin{align*}
		&\phantom{=}(\contr{u}{v_1}-2\cdot\max\{0,\chrg{u}{v_1}\})\cdot w(v_1) \\
		&= (\contr{u}{v_1}-2\cdot\chrg{u}{v_1})\cdot w(v_1)\\&=w^2(u)-w^2(N(u,A)\backslash\{v_1\})-(2\cdot w(u)-w(v_1)-w(N(u,A)\backslash\{v_1\})\cdot w(v_1)\\& \geq (w(u)-w(v_1))^2+(w(v_1)-w(N(u,A)\backslash\{v_1\})\cdot w(N(u,A)\backslash\{v_1\})\\& = (w(u)-w(v_1))^2 + (2\cdot w(v_1)-w(N(u,A)))\cdot w(N(u,A)\backslash\{v_1\}) \\&\geq (1-\sqrt{\epsilon'})\cdot w(v_1)\cdot w(N(u,A)\backslash\{v_1\}).
		\end{align*} As a consequence, \begin{align*}&\phantom{=}\frac{1}{2}\cdot (\contr{u}{v_1}-2\cdot\max\{0,\chrg{u}{v_1}\})\cdot \frac{w(v)}{w(N(u,A)\backslash\{v_1\})}\\&\geq \frac{1-\sqrt{\epsilon'}}{2}\cdot w(v)\\&\stackrel{\eqref{const3}}{\geq} \frac{\epsilon'}{10}\cdot w(v).\end{align*}
	\end{enumerate}
\end{proof}
\begin{lemma} We have
	\[w(A^*)\leq \frac{d}{2}\cdot w(A)-\sum_{v\in A}\sum_{u\in A^*}\mathrm{profit}(u,v)-\sum_{v\in A}(d-1-|N(v,A^*)|)\cdot \frac{w(v)}{2}.\]\label{LemProfit}
	\end{lemma}
\begin{proof}
	Observe that for each $u\in A^*$, at most one of the cases $1$ to $5$ from Definition~\ref{DefProfit} can ever apply for a neighbor $v$ of $u$ because the conditions on $u$ mutually exclude each other. Define $A^*_i$ to be the set of vertices $u\in A^*$ that have neighbors $v\in A$ for which the $i$-th case of the definition of the profit applies, \mbox{$i=1,2,3,4,5$}. Then the sets $A_i^*$, $i=1,2,3,4,5$ are pairwise disjoint. Moreover, for $u\in A^*\backslash\bigcup_{i=1}^5 A^*_i$, $\sum_{v\in A}\mathrm{profit}(u,v)=0$. We now compute the sum $\sum_{v\in A}\mathrm{profit}(u,v)$ for $u\in A^*_i$, \mbox{$i=1,2,3,4,5$}.
	\begin{description}
		\item[$A_1^*$:] By definition, \begin{align*}&\phantom{=}\sum_{v\in A}\mathrm{profit}(u,v)=\sum_{v\in N(u,A)}\mathrm{profit}(u,v)\\&=\sum_{v\in N(u,A)}\left(\frac{1}{2}-\frac{w(u)}{w(N(u,A))}\right)\cdot w(v) \\&= \sum_{v\in N(u,A)}\dfrac{\frac{w(N(u,A))}{2}-w(u)}{w(N(u,A))}\cdot w(v) \\&=\dfrac{\frac{w(N(u,A))}{2}-w(u)}{w(N(u,A))}\cdot w(N(u,A))\\&= \frac{w(N(u,A))}{2}-w(u)=\max\left\{0,\frac{w(N(u,A))}{2}-w(u)\right\}, \end{align*}
		where the last equality follows from the fact that $u$ is $\epsilon'$-payback.
		\item[$A^*_2$:] In this case,
		\begin{align*}
		&\phantom{=}\sum_{v\in A}\mathrm{profit}(u,v)=\sum_{v\in N(u,A)}\mathrm{profit}(u,v)\\& = \sum_{v\in N(u,A)}\frac{(\contr{u}{n(u)}-2\cdot\max\{0,\chrg{u}{n(u)}\})\cdot w(v)}{2\cdot w(N(u,A)) }\\&=\frac{1}{2}\cdot(\contr{u}{n(u)}-2\cdot\max\{0,\chrg{u}{n(u)}\}).
		\end{align*}
		Note that the last term is non-negative by non-negativity of the contribution and Lemma~\ref{LemPropPositiveCharges}.
		\item[$A^*_3$:] Observe that this case only applies for $v\in N(u,A)\backslash\{v_1,v_2\}$, where $v_1=n(u)$ and $v_2=n_2(u)$. Hence, we get
		\begin{align*}
		&\phantom{=}\sum_{v\in A}\mathrm{profit}(u,v)=\sum_{v\in N(u,A)\backslash\{v_1,v_2\}}\mathrm{profit}(u,v)\\&=\sum_{v\in N(u,A)\backslash\{v_1,v_2\}}\dfrac{\frac{w(N(u,A))}{2}-w(u)}{w(N(u,A)\backslash\{v_1,v_2\})}\cdot w(v)\\& = \frac{w(N(u,A))}{2}-w(u) =\max\left\{0,\frac{w(N(u,A))}{2}-w(u)\right\},
		\end{align*}
		where the last equality follows from  our case assumption, which implies \linebreak[4] \mbox{$w(N(u,A))\geq 2\cdot w(u)$}.
		\item[$A^*_4$:] Again, this case only applies for $v\in N(u,A)\backslash\{v_1,v_2\}$, where $v_1=n(u)$ and $v_2=n_2(u)$. Therefore,
		\begin{align*}
		&\phantom{=}\sum_{v\in A}\mathrm{profit}(u,v)=\sum_{v\in N(u,A)\backslash\{v_1,v_2\}}\mathrm{profit}(u,v)\\& = \sum_{v\in N(u,A)\backslash\{v_1,v_2\}}\frac{(\contr{u}{n(u)}-2\cdot\max\{0,\chrg{u}{n(u)}\})\cdot w(v)}{2\cdot w(N(u,A)\backslash\{v_1,v_2\})}\\&=\frac{1}{2}\cdot(\contr{u}{n(u)}-2\cdot\max\{0,\chrg{u}{n(u)}\}).
		\end{align*}
		Note that the last term is non-negative by non-negativity of the contribution and Lemma~\ref{LemPropPositiveCharges}.
		\item[$A^*_5$] This case only applies for $v\in N(u,A)\backslash\{v_1\}$, where $v_1=n(u)$. Therefore,
		\begin{align*}
		&\phantom{=}\sum_{v\in A}\mathrm{profit}(u,v)=\sum_{v\in N(u,A)\backslash\{v_1\}}\mathrm{profit}(u,v)\\& = \sum_{v\in N(u,A)\backslash\{v_1\}}\frac{(\contr{u}{n(u)}-2\cdot\max\{0,\chrg{u}{n(u)}\})\cdot w(v)}{2\cdot w(N(u,A)\backslash\{v_1\})}\\&=\frac{1}{2}\cdot(\contr{u}{n(u)}-2\cdot\max\{0,\chrg{u}{n(u)}\}).
		\end{align*}
		Note that the last term is non-negative by non-negativity of the contribution and Lemma~\ref{LemPropPositiveCharges}.
	\end{description}
Now, we can prove the statement of the lemma: To this end, observe that
\[\sum_{u\in A^*} \frac{w(N(u,A))}{2}=\sum_{v\in A}\frac{|N(v,A^*)|}{2}\cdot w(v) = \frac{d-1}{2}\cdot w(A)-\sum_{v\in A}\frac{d-1-|N(v,A^*)|}{2}\cdot w(v).\]
Furthermore, by Proposition~\ref{PropUpperBoundContr}, Proposition~\ref{PropContrCharge}, and non-negativity of the contribution, we get
\begin{align*} &\phantom{=}\sum_{u\in A^*} \max\{0,\chrg{u}{n(u)}\}\\& = \sum_{u\in A^*}\frac{1}{2}\contr{u}{n(u)}-\sum_{u \in A^*}\frac{1}{2}\cdot(\contr{u}{n(u)}-2\cdot\max\{0,\chrg{u}{n(u)}\})\\& \leq \frac{1}{2}\sum_{u\in A^*}\sum_{v\in A}\contr{u}{v}-\sum_{u \in A^*}\frac{1}{2}\cdot(\contr{u}{n(u)}-2\cdot\max\{0,\chrg{u}{n(u)}\})\\&=\frac{1}{2}\sum_{v\in A}\sum_{u\in A^*}\contr{u}{v} -\sum_{u \in A^*}\frac{1}{2}\cdot(\contr{u}{n(u)}-2\cdot\max\{0,\chrg{u}{n(u)}\})\\&\leq \frac{w(A)}{2}-\sum_{u \in A^*}\frac{1}{2}\cdot(\contr{u}{n(u)}-2\cdot\max\{0,\chrg{u}{n(u)}\})\\&\leq \frac{w(A)}{2}-\sum_{u \in A^*_2\cup A^*_4\cup A^*_5}\frac{1}{2}\cdot(\contr{u}{n(u)}-2\cdot\max\{0,\chrg{u}{n(u)}\}),\end{align*} where the last inequality follows from the fact that \[\contr{u}{n(u)}-2\cdot\max\{0,\chrg{u}{n(u)}\}\geq 0\] by Proposition~\ref{PropContrCharge} and non-negativity of the contribution.
This leads to
\begin{align*} w(A^*) &= \sum_{u\in A^*} w(u) \\
&= \sum_{u\in A^*}\max\left\{w(u),\frac{N(u,A)}{2}\right\}-\max\left\{0,\frac{w(N(u,A))}{2}-w(u)\right\}\\
&=\sum_{u\in A^*} \frac{w(N(u,A))}{2}+\max\left\{0, w(u)-\frac{w(N(u,A))}{2}\right\}\\&\phantom{=}-\max\left\{0,\frac{w(N(u,A))}{2}-w(u)\right\}\\
&=\sum_{u\in A^*} \frac{w(N(u,A))}{2}+\sum_{u\in A^*}\max\{0,\chrg{u}{n(u)}\}\\&\phantom{=}-\sum_{u\in A^*}\max\left\{0,\frac{w(N(u,A))}{2}-w(u)\right\}\\
&=\frac{d-1}{2}\cdot w(A)-\sum_{v\in A}\frac{d-1-|N(v,A^*)|}{2}\cdot w(v)\\&\phantom{=}+\sum_{u\in A^*}\max\{0,\chrg{u}{n(u)}\}\\&\phantom{=}-\sum_{u\in A^*}\max\left\{0,\frac{w(N(u,A))}{2}-w(u)\right\}\\
&\leq \frac{d}{2}\cdot w(A)-\sum_{v\in A}\frac{d-1-|N(v,A^*)|}{2}\cdot w(v)\\&\phantom{=}-\sum_{u \in A^*_2\cup A^*_4\cup A^*_5}\frac{1}{2}\cdot(\contr{u}{n(u)}-2\cdot\max\{0,\chrg{u}{n(u)}\})\\&\phantom{=}-\sum_{u\in A^*_1\cup A^*_3}\max\left\{0,\frac{w(N(u,A))}{2}-w(u)\right\}\\
&=\frac{d}{2}\cdot w(A)-\sum_{v\in A}\frac{d-1-|N(v,A^*)|}{2}\cdot w(v)-\sum_{u\in A^*}\sum_{v\in A}\mathrm{profit}(u,v)\\
&=\frac{d}{2}\cdot w(A)-\sum_{v\in A}\sum_{u\in A^*}\mathrm{profit}(u,v)-\sum_{v\in A}\frac{d-1-|N(v,A^*)|}{2}\cdot w(v).
\end{align*}
\end{proof}
\begin{corollary}
	Let $A_u$ be the set of all vertices $v\in A$ for which the total number of missing or profitable neighbors they have in $A^*$ is larger than $\frac{d-1}{4}$.\\ If $w(A_u)\geq\frac{20}{(d-1)\cdot \epsilon'}\cdot w(A)$, then $w(A^*)\leq \frac{d-1}{2}\cdot w(A)$.\label{CorFewProfitableVertices}
	\end{corollary}
\begin{proof}
	If $v\in A$ has at least $\frac{d-1}{4}$ neighbors in $A^*$ that are either missing, or profitable for $v$, then Lemma~\ref{LemProfitableVertex}, and the definition of missing neighbors, tell us that \[\sum_{u\in A^*} \mathrm{profit}(u,v)+(d-1-|N(v,A^*)|)\cdot\frac{w(v)}{2}\geq \frac{(d-1)\cdot \epsilon'}{40}\cdot w(v).\] By Lemma~\ref{LemProfit}, $w(A_u)\geq \frac{20}{(d-1)\cdot \epsilon'}\cdot w(A)$ would, therefore, imply
	\[w(A^*)\leq \frac{d}{2}\cdot w(A)-\frac{(d-1)\cdot \epsilon'}{40}\cdot w(A_u)\leq \frac{d-1}{2}\cdot w(A).\]
\end{proof}
\begin{lemma}
	Each $u\in A^*$ is one of the following:
	\begin{enumerate}
		\item $\sqrt{\epsilon'}$-single
		\item $\sqrt{\epsilon'}$-double
		\item $\epsilon'$-payback
		\item $\frac{\epsilon'}{2}$-contributive
		\item $\sqrt{2\epsilon'}$-good
	\end{enumerate}
\label{LemTotalClassification}
\end{lemma}
\begin{proof}
	Let $v:=n(u)$. By Lemma~\ref{LemClassificationTv}, each vertex in $T_v$ is $\sqrt{\epsilon'}$-single, $\sqrt{\epsilon'}$-double or $\frac{\epsilon'}{2}$-contributive. So in case $u$ sends positive charges, the lemma is true for $u$. On the other hand, if $u$ does not send positive charges, i.e. $w(N(u,A))\geq 2\cdot w(u)$, then Lemma~\ref{LemClassificationNoCharges} (which is applicable by \eqref{const1} and \eqref{const4}) yields the statement of the lemma since being $\epsilon'$-contributive is a stronger statement than being $\frac{\epsilon'}{2}$-contributive.
\end{proof}
We define the set $\bar{B}\subseteq A$ to contain all vertices $v\in A$ with \[\sum_{u\in T_v}w^2(u)-w^2(N(u,A)\backslash\{v\})\geq \tilde{\epsilon}\cdot w^2(v).\] Note that in particular, all $v$ that possess $u\in T_v$ that is $\sqrt{\epsilon'}$-single fulfill this condition because by Lemma~\ref{LemPropPositiveCharges}, each of the above summands is positive and for $u\in T_v$ $\sqrt{\epsilon'}$-single, we get \begin{align*}\contr{u}{v}\cdot w(v)&=w^2(u)-w^2(N(u,A)\backslash\{v\})\\&\geq (1-\sqrt{\epsilon'})^2\cdot w^2(v)-(\sqrt{\epsilon'}\cdot w(v))^2 \\&= (1-2\sqrt{\epsilon'})\cdot w^2(v)\stackrel{\eqref{const7}}{>}\frac{w^2(v)}{2}\stackrel{\eqref{const0}}{>}\tilde{\epsilon}\cdot w^2(v).\end{align*}
	Our next goal is to show that $w(\bar{B})$ constitutes at most an $\mathcal{O}(\frac{1}{d-1})$-fraction of $w(A)$.
	\begin{lemma}
		If $w(\bar{B})\geq \frac{40}{(d-1)\cdot \epsilon'}\cdot w(A)$, then $w(A^*)\leq \frac{d-1}{2}\cdot w(A)$.\label{LemBbarsmallOrLogImpr}
		\end{lemma}
	\begin{proof}
		Let $A_u$ be defined as in Corollary~\ref{CorFewProfitableVertices}. If $w(A_u)\geq\frac{20}{(d-1)\cdot\epsilon'}\cdot w(A)$, then we are done by Corollary~\ref{CorFewProfitableVertices}. Therefore, assume that $w(A_u)<\frac{20}{(d-1)\cdot\epsilon'}\cdot w(A)$ and let $B:=\bar{B}\backslash A_u$. Then $w(B)> \frac{20}{(d-1)\cdot\epsilon'}\cdot w(A)$. Our goal is to derive the existence of a claw-shaped or circular improvement, proving that this case cannot occur.\\
		Let $v\in B$. By Lemma~\ref{LemTotalClassification}, each $u\in N(v,A^*)$ is $\sqrt{\epsilon'}$-single, $\sqrt{\epsilon'}$-double, $\frac{\epsilon'}{2}$-contributive, $\epsilon'$-payback or $\sqrt{2\epsilon'}$-good. Moreover, the total number of neighbors of $v$ that are missing or profitable for $v$ (cf. Definition~\ref{DefProfitable}) amounts to at most $\frac{d-1}{4}$. Therefore, $v$ possesses at least $\frac{3}{4}\cdot (d-1)$ neighbors $u$ that are $\sqrt{\epsilon'}$-single with $v=n(u)$, $\sqrt{\epsilon'}$-double with $v=n(u)$ or $v=n_2(u)$ or $\sqrt{2\epsilon'}$-good and not $\epsilon'$-payback with $v=n(u)$ or $v=n_2(u)$.\\
		 Next, we show that the number of neighbors of $v$ that are $\sqrt{\epsilon'}$-single with $v=n(u)$ cannot be larger than $\frac{d-1}{4}$. Assume towards a contradiction that this were the case. We have already seen that for a $\sqrt{\epsilon'}$-single vertex $u\in T_v$, $\contr{u}{v}\geq \tilde{\epsilon}\cdot w(v)$. But this means that $v$ would receive a total contribution of $\frac{d-1}{4}\cdot\tilde{\epsilon}\cdot w(v)\stackrel{\eqref{const8}}{>} w(v) $, contradicting the fact that by Proposition~\ref{PropUpperBoundContr}, the total contribution $v$ receives can be at most $w(v)$ because there is no claw-shaped local improvement when LogImp terminates. Thus, $v$ possesses at most $\frac{d-1}{4}$ neighbors that are $\sqrt{\epsilon'}$-single with $v=n(u)$.\\
		As a consequence, $v$ has at least $\frac{3}{4}\cdot (d-1)-\frac{d-1}{4}=\frac{d-1}{2}$ neighbors that are  $\sqrt{\epsilon'}$-double with $v=n(u)$ or $v=n_2(u)$ or $\sqrt{2\epsilon'}$-good with $v=n(u)$ or $v=n_2(u)$, and not $\epsilon'$-payback. \\
		Now, we want to apply Lemma~\ref{LemFindLogarithmicImprovement}. We consider the following graph $G^*$:
		\begin{itemize}
			\item $V(G^*):=A$
			\item $E(G^*):=\{\{n(u),n_2(u)\}:\text{$u$ is $\sqrt{2\epsilon'}$-good and not $\epsilon'$-payback or $\sqrt{\epsilon'}$-double }\wedge\phantom{E(G^*):=\{\}}\{n(u),n_2(u)\}\cap B\neq \emptyset\}$
		\end{itemize}
	The vertex weights are given by $w$. We further set $X:=B$ and $Y:=A\backslash B$.
	By definition of $E(G^*)$, $Y$ is stable and there are no loops at all. By definition of being $\sqrt{2\epsilon'}$-good or $\sqrt{\epsilon'}$-double, we can choose $\alpha:=(1-\sqrt{2\epsilon'})^{-2}$. Moreover, we can pick $t:=\frac{d-1}{2}\geq 5$ by \eqref{const12}. Hence, \[\frac{2\alpha}{t-2}=\frac{4\cdot (1-\sqrt{2\epsilon'})^{-2}}{d-5}\stackrel{\eqref{const9}}{<}\frac{1}{2}.\] This leads to \[\left(1-\frac{2\alpha}{t-2}\right)^{-1}\leq 1+\frac{8\cdot (1-\sqrt{2\epsilon'})^{-2}}{d-5}\] because $\frac{1}{1-x}\leq 1+2x$ for $0\leq x \leq 0.5$. As $Y\subseteq A$, this yields \begin{align*}\left(\left(1-\frac{2\alpha}{t-2}\right)^{-1}-1\right)\cdot w(Y)&\leq \frac{8\cdot (1-\sqrt{2\epsilon'})^{-2}}{d-5}\cdot w(A)<\frac{16}{(d-1)\cdot \epsilon'}\cdot w(A)\\&<\frac{20}{(d-1)\cdot \epsilon'}\cdot w(A) < w(B)=w(X)\end{align*} by \eqref{const10} and \eqref{const11}. If $|V(G)|\leq 1$, our algorithm is optimum. If $|V(G)|\geq 2$ and $|A|=1$, then the unique $v\in A$ must come from $B$ since $w(B)>0$ and it must have an incident loop in $G^*$ and $1\leq 4\cdot\log(|V(G)|)$. Otherwise, we can apply Lemma~\ref{LemFindLogarithmicImprovement} to obtain a cycle $C$ of size at most $4\cdot\log(|A|)\leq 4\cdot\log(|V(G)|)$ and as $Y$ is stable, at least every second vertex of $C$ comes from $B$.\\
	 We need to see how to get a circular local improvement out of $C$. To this end, denote the set of vertices from $A^*$ corresponding to the edges of $C$ by $U_e$, and define
	 \[X:=U_e\cup\bigcup_{v\in V(C)\cap B} T_v.\] Note that \[|X|\leq |E(C)|+(d-1)\cdot |V(C)|=d\cdot |V(C)|\leq 4\cdot d\cdot\log(|V(G)|).\] Let $e=\{n(u),n_2(u)\}\in E(C)$ and let $\{v,z\}:=e$ such that $v\in B$ (and we do not make any assumptions on whether or not $z\in B$). This is possible since every edge of $G^*$ and hence of $C$ must intersect $B$ by definition. We claim that \begin{align}w^2(u)+\frac{1}{2}\cdot w^2(T_v\backslash U_e)>& \frac{w^2(v)+w^2(z)}{2}+w^2(N(u,A)\backslash\{v,z\})\notag\\&+\frac{1}{2}\cdot\sum_{x\in T_v\backslash U_e} w^2(N(x,A)\backslash\{v\}).\label{EqEachEdgeInproving}\end{align}
	We have $\sum_{x\in T_v} w^2(x)-w^2(N(x,A)\backslash\{v\})\geq \tilde{\epsilon}\cdot w^2(v)$ by definition of $B$. Moreover, we know that each vertex $x\in T_v\cap U_e$ sends positive charges and satisfies $n(x)=v$ since it is in $T_v$, and is $\sqrt{\epsilon'}$-double or $\sqrt{2\epsilon'}$-good since it is in $U_e$. As good vertices invoke no positive charges, $x\in T_v\cap U_e$ has to be $\sqrt{\epsilon'}$-double. Hence,
	\begin{align*}w^2(x)-w^2(N(x,A)\backslash\{v\})&\leq w^2(x)-w^2(n_2(x))\\&\leq (1+\sqrt{\epsilon'})^2\cdot w^2(v)-(1-\sqrt{\epsilon'})^2\cdot w^2(v)\\&= 4\sqrt{\epsilon'}\cdot w^2(v).\end{align*}
	Next, as the edge in $E(C)$ that $x\in T_v\cap U_e$ induces is $\{n(x),n_2(x)\}=\{v,n_2(x)\}$, this edge must be incident to $v$. But in the cycle $C$, $v$ has exactly two incident edges. Therefore, $|T_v\cap U_e|\leq 2$ and
	\[\sum_{x\in T_v\backslash U_e} w^2(x)-w^2(N(x,A)\backslash\{v\})\geq (\tilde{\epsilon}-8\sqrt{\epsilon'})\cdot w^2(v).\]
	We have $e=\{n(u),n_2(u)\}=\{v,z\}$, which implies \[w(v)\geq \min\{w(n(u)),w(n_2(u))\}=w(n_2(u)).\]
	As $u$ is $\sqrt{2\epsilon'}$-good or $\sqrt{\epsilon'}$-double, we either get $w(n_2(u))\leq w(n(u))\leq w(u)$ and, hence, \[\frac{w^2(v)+w^2(z)}{2}=\frac{w^2(n(u))+w^2(n_2(u))}{2}\leq w^2(u),\] or   \[w(u)<w(n(u))\leq\frac{1}{1-\sqrt{2\epsilon'}}\cdot w(u)\stackrel{\eqref{const1}}{\leq} (1+2\sqrt{2\epsilon'})\cdot w(u)\] and \[(1-\sqrt{2\epsilon'})\cdot w(u)<(1-\sqrt{2\epsilon'})\cdot w(n(u))\leq w(n_2(u))\leq w(n(u)),\] leading to \begin{align*}\frac{w(v)^2+w(z)^2}{2}&=\frac{w(n(u))^2+w(n_2(u))^2}{2}\leq \frac{2+8\cdot\sqrt{2\epsilon'}+16\epsilon'}{2}\cdot w^2(u)\\&= w^2(u)+(4\sqrt{2\epsilon'}+8\epsilon')\cdot w^2(u)\\&\leq w^2(u)+\frac{4\sqrt{2\epsilon'}+8\epsilon'}{(1-\sqrt{2\epsilon'})^2}\cdot w^2(v).\end{align*} Hence, the weaker inequality\[\frac{w(v)^2+w(z)^2}{2}\leq  w^2(u)+\frac{4\sqrt{2\epsilon'}+8\epsilon'}{(1-\sqrt{2\epsilon'})^2}\cdot w^2(v)\] holds in either case.  Additionally, as $w(v)\geq w(n_2(u))\geq (1-\sqrt{2\epsilon'})^2\cdot w(u)$ and $w(N(u,A))\leq (2+\epsilon')\cdot w(u)$ by definition of $\sqrt{\epsilon'}$-double or $\sqrt{2\epsilon'}$-good, and not $\epsilon'$-payback, we get \begin{align*}w^2(N(u,A)\backslash\{v,z\})&=w^2(N(u,A)\backslash\{n(u),n_2(u)\})\\&\leq (w(N(u,A))-w(n(u))-w(n_2(u)))^2\\&\leq(2+\epsilon'-(1-\sqrt{2\epsilon'})-(1-\sqrt{2\epsilon'})^2)^2\cdot w^2(u)\\&<(3\sqrt{2\epsilon'})^2\cdot w^2(u)\\&=18\epsilon'\cdot w^2(u)\\&\leq 18\epsilon'\cdot \frac{w^2(v)}{(1-\sqrt{2\epsilon'})^4}\end{align*}
	This gives
	\begin{align*}
	&\phantom{=}w^2(u)+\frac{1}{2}\sum_{x\in T_v\backslash U_e} w^2(x)-w^2(N(x,A)\backslash\{v\}) \\&\geq w^2(u)+\frac{\tilde{\epsilon}}{2}\cdot w^2(v)-4\sqrt{\epsilon'}\cdot w^2(v)\\ &\geq \frac{w^2(v)+w^2(z)}{2}+w^2(N(u,A)\backslash\{v,z\})+\frac{\tilde{\epsilon}}{2}\cdot w^2(v)\\&\phantom{=}-\left(4\sqrt{\epsilon'}+\frac{4\sqrt{2\epsilon'}+8\epsilon'}{(1-\sqrt{2\epsilon'})^2}+\frac{18\epsilon'}{(1-\sqrt{2\epsilon'})^4}\right)\cdot w^2(v)\\&>\frac{w^2(v)+w^2(z)}{2}+w^2(N(u,A)\backslash\{v,z\}),
	\end{align*} where the last strict inequality follows from \eqref{const5} and $w(v)>0$ since weights are positive. This proves our claim \eqref{EqEachEdgeInproving}.\\
	Now, when summing up all of the inequalities \eqref{EqEachEdgeInproving} for $e\in E(C)$, each $v\in V(C)$ appears exactly twice (once per incident edge) and each $u\in U_e$ occurs exactly once. Moreover, for each $v_0\in B$, the terms for $T_{v_0}$ appear at most twice since $v_0$ has only two incident edges (but they need not appear at all if for both incident edges the other vertex $v_1$ is from $B$ as well and we pick it as $v$ and $v_0$ as $z$). However, we have seen that 
	\begin{equation}\frac{1}{2}\cdot\sum_{x\in T_v\backslash U_e} w^2(x)-w^2(N(x,A)\backslash\{v\})\geq \frac{1}{2}\cdot (\tilde{\epsilon}-8\sqrt{\epsilon'})\cdot w^2(v)\geq 0,\label{EqImprovementFromTv}\end{equation} where the last inequality follows from \eqref{const13}. Hence, by summing up all the inequalities \eqref{EqEachEdgeInproving} for the edges and the latter inequalities \eqref{EqImprovementFromTv} in a way that each set $T_v$ for $v\in V(C)\cap B$ appears exactly twice in total, we get, as the sets $T_v\subseteq\{u:n(u)=v\}$ are pairwise disjoint,
	\begin{align*}
	w^2(X)&=w^2(U_e\cup\bigcup_{v\in V(C)\cap B}T_v)\\&=\sum_{u\in U_e} w^2(u)+\sum_{v\in V(C)\cap B}\sum_{x\in T_v\backslash U_e} w^2(x) \\&> w^2(V(C))+\sum_{u\in U_e} w^2(N(u,A)\backslash\{n(u),n_2(u)\})\\&\phantom{=}+\sum_{v\in V(C)\cap B}\sum_{x\in T_v\backslash U_e} w^2(N(x,A)\backslash\{v\})\\
	&\geq w^2(N(U_e\cup\bigcup_{v\in V(C)\cap B} T_v, A))=w^2(N(X,A)),\end{align*}
	showing that we obtain a local improvement. Moreover, we have \[|U_e|=|E(C)|\leq 4\cdot \log(|V(G)|)\] and $|T_v|\leq d-1$ by $d$-claw freeness of $G$ and since $T_v=\{v\}$ if $v\in A^*\cap A$. Hence, choosing $U:=U_e$ and $Y_v:=T_v\backslash U_e$ for $v\in V(C)\cap B$ and $Y_v:=\emptyset$ for $v\in V(C)\backslash B$, we see that our local improvement is circular by \eqref{EqEachEdgeInproving} and \eqref{EqImprovementFromTv}, provided it is a subset of $V\backslash A$. To this end, note that $V(C)\subseteq A\backslash A^*$ since vertices in $A\cap A^*$ are only adjacent to themselves in $A^*$, and, hence, isolated in $G^*$. As a consequence, as all vertices in $X=U_e\cup\bigcup_{v\in V(C)\cap B} T_v$ are adjacent to some $v\in V(C)\subseteq A\backslash A^*$, they cannot be contained in $A^*\cap A$ either, because in this case, they could only be adjacent to themselves in $A$ and in particular not to a vertex in $A\backslash A^*$. Hence, $X\subseteq A^*\backslash A\subseteq V\backslash A$ defines a circular improvement and we have arrived at the desired contradiction.
	\end{proof}
We want to show that for $v\in A\backslash \bar{B}$, the total charges send to $v$ are small. 
\begin{lemma}
	Let $v\in A\backslash \bar{B}$. Then $\sum_{u\in T_v}\chrg{u}{v}\leq \frac{\tilde{\epsilon}}{2}\cdot w(v)$.\label{LemSmallChargesNotBbar}
	\end{lemma}
\begin{proof}
	By Lemma~\ref{LemPropPositiveCharges}, we get \[\sum_{u\in T_v}\chrg{u}{v}\cdot w(v)\leq \frac{1}{2}\cdot\sum_{u\in T_v} w^2(u)-w^2(N(u,A)\backslash\{v\}).\] As $v\in A\backslash\bar{B}$, we must have $\sum_{v\in T_v}w^2(u)-w^2(N(u,A)\backslash\{v\})\leq\tilde{\epsilon}\cdot w^2(v)$, implying the claim.
\end{proof}
Finally, we have all parts together to prove Theorem~\ref{TheoMainResult}. For the sake of readability, we restate it once again.
\begin{RestateTheoMainResult}
	For any $\delta >0$, there exists $d_\delta\in\mathbb{N}$ with the following property:
	If $A^*$ is an optimum solution to the MWIS in a $d$-claw free graph $G$ for some $d\geq d_\delta$ and $A$ denotes the solution returned by LogImp, then we have \[w(A^*)\leq \frac{d-1+\delta}{2}\cdot w(A).\]
\end{RestateTheoMainResult}
\begin{proof}
	By Lemma~\ref{LemBbarsmallOrLogImpr}, we can assume that $w(\bar{B})< \frac{40}{(d-1)\cdot \epsilon'}\cdot w(A)$ since we are done otherwise.
	By \eqref{const6}, it suffices to show that
	\[w(A^*)\leq \frac{d-1}{2}\cdot w(A)+\frac{\tilde{\epsilon}}{2}\cdot w(A)+\frac{20}{(d-1)\cdot \epsilon'}\cdot w(A).\]
	We know that when LogImp terminates, there is no more local improvement and in particular, $A$ is a maximal independent set in $G$.
	Hence, Theorem~\ref{TheoApproxFactor} and Lemma~\ref{LemWeightsNeighborhoodsdminus1} tell us that
	\begin{align*}w(A^*)&\leq \frac{d-1}{2}\cdot w(A)+\sum_{u\in A^*:\chrg{u}{n(u)}>0} \chrg{u}{n(u)}\\&=\frac{d-1}{2}\cdot w(A)+\sum_{v\in A}\sum_{u\in T_v} \chrg{u}{v}.\end{align*}
	In addition to that, we know that by Lemma~\ref{LemBoundCharges}, vertices $v\in \bar{B}\subseteq A$ receive total charges of at most $\frac{w(v)}{2}$, whereas vertices in $A\backslash \bar{B}$ receive total charges of at most $\frac{\tilde{\epsilon}}{2}\cdot w(v)$ by Lemma~\ref{LemSmallChargesNotBbar}. This leads to 
	\[w(A^*)\leq \frac{d-1}{2}\cdot w(A)+\frac{w(\bar{B})}{2}+\frac{\tilde{\epsilon}}{2}\cdot w(A\backslash\bar{B})\leq \frac{d-1}{2}\cdot w(A)+\frac{\tilde{\epsilon}}{2}\cdot w(A)+\frac{w(\bar{B})}{2}.\] Hence, the fact that $w(\bar{B})< \frac{40}{(d-1)\cdot \epsilon'}\cdot w(A)$ allows us to conclude that
	\[w(A^*)\leq \frac{d-1}{2}\cdot w(A)+\frac{\tilde{\epsilon}}{2}\cdot w(A)+\frac{20}{(d-1)\cdot \epsilon'}\cdot w(A)\] as desired. 
\end{proof}
As we have already seen how to derive Theorem~\ref{TheoSequenceEpsilons} from Theorem~\ref{TheoMainResult}, this concludes the analysis of LogImp.
\section{Achieving a polynomial runtime for weighted $k$-Set Packing \label{SecPolyTime}}
The previous considerations result in a quasi-polynomial running time bound for each iteration of LogImp (and we have seen how to ensure a polynomial number of iterations by scaling and truncating the weight function). While it is unclear how to get down to a polynomial running time for the general MWIS in $d$-claw free graphs, for the $k$-Set Packing Problem, which is the main application we have in mind, a polynomial running time can be achieved by means of color coding similar as in \cite{FurerYu} (but in a much simpler way). Recall that the conflict graph $G=(V,E)$ of a $k$-Set Packing instance is $d:=k+1$-claw free. First, note that we have already seen how to find claw-shaped local improvements in polynomial time since they can have at most $d-1=k$ elements. It, therefore, remains to see how we can check for circular local improvements. To this end, observe that each independent set $X\subseteq V\backslash A$ for which $U\subseteq X\cap\{u\in V\backslash A: |N(u,A)|\geq 2\}$ of cardinality at most $4\cdot \log(|V|)$ satisfying conditions (i) to (iii) from Definition~\ref{DefCircular} exists, automatically fulfills $w^2(N(X,A))< w^2(X)$ because
\begin{align*}
w^2(X) &= w^2(U)+\sum_{v\in V(C)} w^2(Y_v) = \sum_{u\in U} w^2(u)+\frac{w^2(Y_{n(u)}\cup Y_{n_2(u)})}{2} \\&> \sum_{u\in U} \frac{w^2(n(u))+w^2(n_2(u))}{2}+w^2(N(u,A)\backslash\{n(u), n_2(u)\})\\&\phantom{=}+\frac{1}{2}\cdot\sum_{u\in U}\left(\sum_{x\in Y_{n(u)}} w^2(N(x,A)\backslash\{n(u)\})+\sum_{x\in Y_{n_2(u)}} w^2(N(x,A)\backslash\{n_2(u)\})\right)\\
&\stackrel{(*)}{=}\sum_{v\in V(C)} w^2(v)+\sum_{x\in Y_v} w^2(N(x,A)\backslash\{v\})+\sum_{u\in U} w^2(N(u,A)\backslash\{n(u),n_2(u)\}))\\
&\geq \sum_{v\in V(C)} w^2(v)+w^2(N(Y_v,A)\backslash\{v\})+\sum_{u\in U} w^2(N(u,A)\backslash V(C))\\
&\geq w^2(N(X,A)).
\end{align*}
Here, the equation labeled $(*)$ follows from the fact that \[C=\left(\bigcup_{u\in U} \{n(u),n_2(u)\}, \{e_u=\{n(u),n_2(u)\}, u\in U\}\right)\] forms a cycle, meaning that each $v\in V(C)$ occurs exactly twice among all of the sets $\{n(u),n_2(u)\},u\in U$.
Hence, any independent set $X\subseteq V\backslash A$, for which a subset $U\subseteq X\cap\{u\in V\backslash A: |N(u,A)|\geq 2\}$ of cardinality at most $4\cdot \log(|V|)$ subject to items (i) to (iii) from Definition~\ref{DefCircular} exists, constitutes a circular local improvement.
Recall that the in the reduction from $k$-Set Packing to the MWIS in $d:=k+1$-claw free graphs, each vertex corresponds to a given set $S\in\mathcal{S}$, and edges correspond to non-empty set intersections. To simplify notation, we identify a set $S\in\mathcal{S}$ and its corresponding vertex in the conflict graph \[G_{\mathcal{S}}=(\mathcal{S},\{\{S_1,S_2\}:S_1,S_2\in\mathcal{S}, S_1\neq S_2, S_1\cap S_2\neq \emptyset\}).\] Moreover, we might just write $G=(V,E)$ to refer to $G_{\mathcal{S}}$. \\
In order to search for a circular improvement, we first compute maps $n$ and $n_2$ meeting the requirements of Definition~\ref{DefCircular} and such that the domain of $n$ is $V\backslash A$. These maps are guaranteed to exist if we have made sure that there is no claw-shaped improvement, and can be easily determined in polynomial time. Then, we define an auxiliary multi-graph $H$ as follows:\\ The vertices of $H$ are pairs $(v,Y)$, where $v\in A$ and $Y\subseteq \{u\in V\backslash A: n(u)=v\}$ is an independent set of cardinality at most $k=d-1$. Then for each $v$, there are at most $\sum_{i=0}^{d-1}|V|^i =\frac{|V|^d-1}{|V|-1}< |V|^d$ many possible choices for $Y$, so $|V(H)| < |V|^{d+1}$.\\ As far as the set of edges is concerned, for each $u\in \{x\in V\backslash A: |N(x,A)|\geq 2\}$ and each pair of vertices $(n(u), Y_{n(u)}),(n_2(u), Y_{n_2(u)})\in V(H)$ such that \begin{align*}
w^2(u)+\frac{w^2(Y_{n(u)}\cup Y_{n_2(u)})}{2}&> \frac{w^2(n(u))+w^2(n_2(u))}{2}+w^2(N(u,A)\backslash\{n(u), n_2(u)\})\\&\phantom{=}+\frac{1}{2}\cdot\sum_{x\in Y_{n(u)}} w^2(N(x,A)\backslash\{n(u)\})\\&\phantom{=}+\frac{1}{2}\cdot\sum_{x\in Y_{n_2(u)}} w^2(N(x,A)\backslash\{n_2(u)\}),\end{align*} and $\{u\}$, $Y_{n(u)}$ and $Y_{n_2(u)}$ are pairwise disjoint and their union is independent,
we add an edge $\{(n(u),Y_{n(u)}), (n_2(u), Y_{n_2(u)})\}$ to $E(H)$. We say that this edge is \emph{induced by $u$}. As we have as most $|V|$ many possible choices for $u$ and at most $|V|^d$ possibilities to choose each of $Y_{n(u)}$ and $Y_{n_2(u)}$, we get $|E(H)|\leq |V|^{2d+1}=|V|^{2k+3}$.\\ For convenience, let $\mu:E(H)\rightarrow V\backslash A$ such that $e\in E(H)$ is induced by the vertex $\mu(e)$.  Now, by definition, there is a one-to-one correspondence between cycles $C$ of length at most $4\cdot\log |V|$ such that $\{\mu(e),e\in E(C)\}\cup\bigcup_{(v,Y)\in V(C)}Y$ defines an independent set a.k.a.\ a disjoint sub-family of $\mathcal{S}$, and circular improvements. Hence, it remains to see how to find such a cycle, or decide that none exists, in polynomial time.\\ First, for each pair of parallel edges, we can check in polynomial time whether or not it yields a circular improvement, so we can restrict ourselves to cycles of length at least $3$ (and at most $4\cdot \log(|V|)$) in the following.\\ To find these, we want to apply the \emph{color coding} technique. For this purpose, we introduce the following terminology:
\begin{definition}[$t$-perfect family of hash functions, \cite{ColorCoding}]
	For $t,m\in\mathbb{N}$ with $t\leq m$, a family $\mathcal{F}\subseteq {}^{\{1,\dots,m\}}\{1,\dots,t\}$ of functions mapping $\{1,\dots,m\}$ to $\{1,\dots,t\}$ is called a \emph{$t$-perfect family of hash functions} if for all $I\subseteq \{1,\dots,m\}$ of size at most $t$, there is $f\in\mathcal{F}$ with $f\upharpoonright I$ injective.
	\end{definition}
\begin{theorem}[stated in \cite{ColorCoding} referring to \cite{SchmidtSiegel}]
For $t,m\in\mathbb{N}$ with $t\leq m$, a $t$-perfect family $\mathcal{F}$ of hash functions of cardinality $\mathcal{O}(2^{\mathcal{O}(t)}\cdot (\log (m))^2)$, where each function is encoded using $\mathcal{O}(t)+2\log\log m$ many bits, can be explicitly constructed such that the query time is constant.
\end{theorem}
For our application, let $\mathcal{F}$ be a $t:=4\cdot (k+1)\cdot k\cdot\log(|\mathcal{S}|)=4\cdot (k+1)\cdot k\cdot\log(|V|)$-perfect family of functions with domain $\bigcup\mathcal{S}$, i.e.\ the underlying universe of the $k$-Set Packing Problem (or, more precisely, its restriction to the set of elements that appear in at least one set). Clearly, $|\bigcup\mathcal{S}|\leq k\cdot |\mathcal{S}|$, so we obtain such a family $\mathcal{F}$ of size  \[\mathcal{O}(2^{\mathcal{O}(k^2\cdot\log(|\mathcal{S}|))}\cdot (\log (k\cdot |\mathcal{S}|))^2)=\mathcal{O}(|\mathcal{S}|^{\mathcal{O}(k^2)}\cdot (\log (k\cdot |\mathcal{S}|))^2),\] which is polynomial. For each $f\in\mathcal{F}$, we do the following:\\ We assign to an edge $e=\{(v_1,Y_1),(v_2,Y_2)\}\in E(H)$ the set of colors \[\mathrm{col}(e):=f(\mu(e))=\{f(x),x\in \mu(e)\},\] where we interpret $\mu(e)$ as the corresponding $k$-set. For a vertex $v=(z,Y)\in V(H)$, we define \[\mathrm{col}(v):=f(\bigcup Y)=\{f(x):\exists S\in Y: x\in S\},\] i.e.\ we color $v$ with all of the colors occurring among the elements of the $k$-sets in $Y$. Following \cite{FurerYu}, we call a path or cycle $P$ in $H$ \emph{colorful} if the color sets assigned to its edges and vertices are pairwise disjoint.\\
 We apply dynamic programming to seek for a colorful cycle of length at most $4\cdot\log|V|=4\log|\mathcal{S}|$. To do so, we compute the Boolean values $\mathrm{Path}(v,w,C,i)$, where $v, w\in V(H)$, $C\subseteq \{1,\dots,t\}$ and $i\in\{0,\dots,4\cdot\log|V|-1\}$, telling us whether there is a colorful $v$-$w$-path with $i$ edges in $H$ such that the union of the color sets of its vertices and edges is $C$. We have to compute at most \begin{align*}|V(H)|^2\cdot 2^t\cdot 4\cdot\log|V|&\leq|V|^{2(d+1)}\cdot 2^{4\cdot k\cdot(k+1)\cdot\log(|V|)}\cdot4\cdot\log|V|\\& = |V|^{4k^2+6k+4}\cdot4\cdot\log|V|\end{align*} of these values, which is polynomially bounded. Whenever we set a value $\mathrm{Path}(v,w,C,i)$ with $i\geq 2$ to true (correctly), we check whether there is an edge $e=\{v,w\}$ such that $\mathrm{col}(e)\cap C=\emptyset$, i.e.\ an edge that completes the current path to a colorful cycle. In this case, we can use backlinks (which we do not mention explicitly in Algorithm~\ref{AlgDynProg} for the sake of readability) to retrace the corresponding colorful cycle.\\
\begin{algorithm}[t]
	\DontPrintSemicolon
	\KwIn{$H$, $f$}
	\KwOut{a colorful cycle of length at least $3$ and at most $4\cdot\log(|V|)$ or the information that none exists}
	Delete all edges $e=\{v,w\}$ such that $\mathrm{col}(e)$, $\mathrm{col}(v)$ and $\mathrm{col}(w)$ are not pairwise disjoint.\label{LineDeleteEdges}\;
	\ForEach{$v,w\in V(H)$, $C\subseteq \{1,\dots,t\}$, $i\in\{0,\dots,4\cdot\log|V|-1\}$}{$\mathrm{Path}(v,w,C,0)\gets 0$\;}
	\ForEach{$v\in V(H)$}{
		$\mathrm{Path}(v,v,\mathrm{col}(v),0)\gets 1$\;
	}
	\For{$i\gets 1$ to $4\cdot\log(|V|)-1$}{
		\ForEach{$C\subseteq\{1,\dots,t\}$}{
			\ForEach{$s\in V(H)$}{
				\ForEach{$t\in V(H)$}{
					\ForEach{$e=\{s,v\}\in \delta_H(s)$}{
						\If{$\mathrm{col}(e)\cup \mathrm{col}(s)\subseteq C$}{
							$\mathrm{Path}(s,t,C, i)\gets \mathrm{Path}(s,t,C, i)\vee \mathrm{Path}(v,t,C\backslash (\mathrm{col}(e)\cup\mathrm{col}(s)),i-1)$\label{LineUpdate}\;}}
					\If{$i\geq 2\wedge\mathrm{Path}(s,t,C,i)$}
					{\ForEach{$e=\{s,t\}\in E(H)$}{
							\If{$\mathrm{col}(e)\cap C = \emptyset$}{Use backlinks to recover the edges of a colorful $s$-$t$-path $P$ of length $i$ with colors from $C$.\;
								\textbf{return} $P+e$\;}}}}}}}
	\textbf{return} "No colorful cycle found"	
	\caption{Dynamic program searching for a colorful cycle.}\label{AlgDynProg}
\end{algorithm}
 Consider Algorithm~\ref{AlgDynProg}, which we employ to compute the values $\mathrm{Path}(v,w,C,i)$ and seek for a colorful cycle. Its running time is polynomial since the size of $H$ (w.r.t. the size of $G$ or $\mathcal{S}$, respectively) and $2^t=|V|^{4\cdot k\cdot(k+1)}$ are.\\ Correctness follows by induction on $i$. For $i=0$, the initialization is correct because the only paths of length zero consist of precisely one vertex $v\in V(H)$ and zero edges and feature exactly the colors of the respective vertex.\\ If for some $s,t\in V(H)$, there exists a colorful $s$-$t$-path with color set $C$ and $i>0$ edges, then the edge set of the path is non-empty and must, therefore, contain an edge $e\in\delta_H(s)$, and we have $\mathrm{col}(s)\cup\mathrm{col}(e)\subseteq C$. Let $v$ be the other endpoint of $e$. Then $P-e$ constitutes a colorful $v$-$t$ path of length $i-1$ with color set $C\backslash(\mathrm{col}(e)\cup\mathrm{col}(s))$. By the induction hypothesis, $\mathrm{Path}(v,t,C\backslash (\mathrm{col}(e)\cup\mathrm{col}(s)),i-1)=1$, and $\mathrm{Path}(s,t,C,i)$ is set to $1$ in the update step (line~\ref{LineUpdate}). Conversely, if we perform an update from $0$ to $1$ in line~\ref{LineUpdate}, then $\mathrm{col}(e)\cup\mathrm{col}(s)\subseteq C$ and by the induction hypothesis, there exists a colorful $v$-$t$-path with colors from $C\backslash (\mathrm{col}(e)\cup\mathrm{col}(s))$ and $i-1$ edges. In particular, $e$ and $s$ and their colors have not occurred on this path yet. Moreover, line~\ref{LineDeleteEdges} ensures that the color sets of $e$ and $s$ are disjoint. So the update step is correct and by induction, all of the $\mathrm{Path}$-values are set correctly.\\ Now, there exists a colorful cycle of length at least $3$ and at most $4\cdot\log|V|$ if and only if for some $s\neq t$, $C$ and $2\leq i \leq 4\cdot\log|V|-1$, we have $\mathrm{Path}(s,t,C,i)=1$ and there exists an edge $e=\{s,t\}$ with $\mathrm{col}(e)\cap C=\emptyset$. Hence, Algorithm~\ref{AlgDynProg} is correct and we have found a polynomial time algorithm to check whether a colorful cycle of length at least $3$ and at most $4\cdot\log(|V|)$ exists, and return one if this is the case.\\ For a colorful cycle $C$, the sets of colors assigned to the (union of) $k$-sets in $Y$, $(x,Y)\in V(C)$ and the colors assigned to the $k$-sets $\mu(e)$, $e\in E(C)$ are pairwise distinct. In particular, $k$-sets from different sets $Y$, $k$-sets corresponding to different edges and $k$-sets from sets $Y$ and $k$-sets corresponding to edges are pairwise distinct. Moreover, when constructing $H$, we made sure that for each $(v,Y)\in V(H)$, $Y$ is independent, meaning that the contained $k$-sets are pairwise distinct. Hence, $\bigcup_{(x,Y)\in V(C)} Y\cup\{\mu(e), e\in E(C)\}$ defines a circular improvement. \\ On the other hand, if $X$ defines a circular improvement, then \[|X|\leq 4\cdot d\cdot\log(|V|)=4\cdot (k+1)\cdot \log(|V|),\] the sets in $X$ are pairwise disjoint and \[|\bigcup X| \leq 4\cdot (k+1)\cdot k\cdot \log(|V|)=t.\] Hence, there exists $f\in\mathcal{F}$ assigning distinct colors to all elements of $\bigcup X$ and we find a circular improvement in this case. Consequently, we have found a polynomial time method to check whether a circular improvement exists and find one, if this is the case. This shows that in the context of $k$-Set Packing, we can implement each iteration of LogImp to run in polynomial time, and a polynomial number of iterations can be ensured by scaling and truncating the weight function.
\FloatBarrier
\section{Searching for local improvements of logarithmic size cannot result in an approximation guarantee better than $\frac{d-1}{2}$\label{SecExamples}}
In this section, we show that we cannot hope to get beyond an approximation guarantee of $\frac{d-1}{2}$ by considering local improvements of $w^\alpha$ for some $\alpha\in\mathbb{R}$ of at most logarithmic size. More precisely, we prove the following theorem:
\begin{theorem}
	Let $d\geq 3$, $\alpha\in\mathbb{R}$, $0<\epsilon <1$ and $C>0$. Then for each $N_0\in\mathbb{N}$, there exists an undirected, simple graph $G$ with $|V(G)|\geq N_0$ vertices, a weight function $w:V(G)\rightarrow\mathbb{R}_{>0}$, and an independent set $A\subseteq V(G)$ with the following properties:
	\begin{romanenumerate}
		\item Each vertex in $G$ has degree at least $2$ and at most $d-1$. In particular, $G$ is $d$-claw free.
		\item For each independent $X\subseteq V(G)$ with $|X|\leq C\cdot\log(|V(G)|)$, we have\\ $w^\alpha(X)\leq w^\alpha(N(X,A))$.\\ (This means that there is no local improvement of $w^\alpha(A)$ of size at most $C\cdot\log(|V(G)|)$.)
		\item For each optimum solution $A^*$, we have $w(A^*)\geq \frac{d-1-\epsilon}{2}\cdot w(A)$.
		\end{romanenumerate}\label{TheoNoBetterThanDMinus1Half}
\end{theorem}
\begin{algorithm}[t]
	\DontPrintSemicolon
	\KwIn{a $d$-claw free graph $G$,\; a (positive) weight function $w:V(G)\rightarrow\mathbb{R}_{>0}$},\; parameters $\alpha\in\mathbb{R}$ and $C>0$\;
	\KwOut{an independent set $A\subseteq V(G)$}\;
	$A\gets\emptyset$\;
	\While{$\exists$ inpendent set $X\subseteq V(G)$: $|X|\leq C\cdot \log(|V(G)|)$ $\wedge$ $w^\alpha(N(X,A)) < w^\alpha(X)$}{
		$A\gets A\backslash N(X,A)\cup X$\;}
	\textbf{return} $A$\;
	\caption{Parametrized local improvement algorithm for the MWIS}\label{AlgParametrized}
\end{algorithm}
\begin{algorithm}[t]
	\DontPrintSemicolon
	\KwIn{a collection $\mathcal{S}$ of sets each of cardinality at most $k$,\; a (positive) weight function $w:\mathcal{S}\rightarrow\mathbb{R}_{>0}$},\; parameters $\alpha\in\mathbb{R}$ and $C>0$\;
	\KwOut{a sub-collection $A\subseteq \mathcal{S}$ consisting of pairwise disjoint sets}\;
	$A\gets\emptyset$\;
	\While{$\exists$ a collection of pairwise disjoint sets $X\subseteq \mathcal{S}$: $|X|\leq C\cdot \log(|\mathcal{S}|)$ $\wedge$ $w^\alpha(\{v\in A: \exists u\in X: u\cap v\neq \emptyset\}) < w^\alpha(X)$}{
		$A\gets A\backslash \{v\in A: \exists u\in X: u\cap v\neq \emptyset\}\cup X$\;}
	\textbf{return} $A$\;
	\caption{Parametrized local improvement algorithm for the $k$-Set Packing Problem}\label{AlgParametrizedSetPacking}
\end{algorithm}
Before we engage in the proof of the theorem, we would like to point out that it implies the following statements:
\begin{corollary}
	Let $d\geq 3$, $\alpha\in\mathbb{R}$, $0<\epsilon <1$ and $C>0$. Then for each $N_0\in\mathbb{N}$, there exists an instance $(G,w)$ of the MWIS in $d$-claw free graphs on $|V(G)|\geq N_0$ vertices such that Algorithm~\ref{AlgParametrized} does not yield a better approximation ratio than $\frac{d-1-\epsilon}{2}$.
\end{corollary}
\begin{proof}
	By Theorem~\ref{TheoNoBetterThanDMinus1Half}, there exists an undirected graph $G$ with $|V(G)|\geq N_0$ vertices, a weight function $w:V(G)\rightarrow\mathbb{R}_{>0}$, and an independent set $A'\subseteq V(G)$ with the following properties:
	\begin{romanenumerate}
		\item Each vertex in $G$ has degree at least $2$ and at most $d-1$. In particular, $G$ is $d$-claw free.
		\item For each independent set $X\subseteq V(G)$ with $|X|\leq C\cdot\log(|V(G)|)$, we have\\ $w^\alpha(X)\leq w^\alpha(N(X,A'))$.\\ (This means that there is no local improvement of $w^\alpha(A')$ of size at most $C\cdot\log(|V(G)|)$.)
		\item For each optimum solution $A^*$, we have $w(A^*)\geq \frac{d-1-\epsilon}{2}\cdot w(A')$.
	\end{romanenumerate}
	Now, as $w>0$, Algorithm~\ref{AlgParametrized} may just pick $A'$ vertex by vertex and then return the found solution. Hence, the approximation ratio of Algorithm~\ref{AlgParametrized} on the given instance is no better than $\frac{d-1-\epsilon}{2}$.
\end{proof}
\begin{corollary}
	Let $k\geq 3$, $\alpha\in\mathbb{R}$, $0<\epsilon <1$ and $C>0$. Then for each $N_0\in\mathbb{N}$, there exists an instance $(\mathcal{S}, w)$ of weighted $k$-Set Packing Problem with $|\mathcal{S}|\geq N_0$, such that Algorithm~\ref{AlgParametrizedSetPacking} does not yield a better approximation ratio than $\frac{k-\epsilon}{2}$. \label{CorInstancesSetPacking}
\end{corollary}
\begin{proof}
	Note that Algorithm~\ref{AlgParametrizedSetPacking} is just Algorithm~\ref{AlgParametrized}, applied to the conflict graph of the given instance of the weighted $k$-Set Packing Problem.
	Again, by Theorem~\ref{TheoNoBetterThanDMinus1Half}, there exists an undirected, simple graph $G$ with $|V(G)|\geq N_0$ vertices, a weight function $w':V(G)\rightarrow\mathbb{R}_{>0}$, and an independent set $A'\subseteq V(G)$ with the following properties:
	\begin{romanenumerate}
		\item Each vertex in $G$ has degree at least $2$ and at most $k$.
		\item For each independent set $X\subseteq V(G)$ with $|X|\leq C\cdot\log(|V(G)|)$, we have\\ $w'^\alpha(X)\leq w'^\alpha(N(X,A'))$.\\ (This means that there is no local improvement of $w'^\alpha(A')$ of size at most $C\cdot\log(|V(G)|)$.)
		\item For each optimum solution $A'^*$, we have $w'(A'^*)\geq \frac{k-\epsilon}{2}\cdot w'(A')$.
	\end{romanenumerate}
	Define $\mathcal{S}:=\{\delta(v): v\in V(G)\}$, where $\delta(v)$ denotes the set of incident edges of $v$. Then each of the sets in $\mathcal{S}$ is of cardinality at most $k$. Moreover, note that for two distinct vertices $v,w\in V(G)$, the sets $\delta(v)$ and $\delta(w)$ have to be distinct because both $v$ and $w$ have degree at least $2$, but there cannot be two edges between $v$ and $w$ since $G$ is simple. We can, hence, define $w:\mathcal{S}\rightarrow\mathbb{R}_{>0}$ by setting $w(\delta(v)):=w'(v)$ for $v\in V(G)$. As we have a one-to-one correspondence between independent sets in $G$ and sub-families of $\mathcal{S}$ the sets of which are pairwise disjoint (mapping $B\subseteq V(G)$ to $\{\delta(v), v\in B\}$ and vice versa), we know that there is no sub-collection $X$ consisting of at most $C\cdot\log|\mathcal{S}|$ pairwise disjoint sets such that $X$ improves $w^\alpha(\{\delta(v):v\in A'\})$ and additionally, we have $w(A^*)\geq \frac{k-\epsilon}{2}\cdot w(\{\delta(v):v\in A'\})$ for each optimum solution $A^*$ to the weighted $k$-Set Packing Problem. As Algorithm~\ref{AlgParametrizedSetPacking} may just pick $\{\delta(v):v\in A'\}$ set by set, the claim follows.
\end{proof}
\begin{corollary}
	For any $\alpha\in\mathbb{R}$, $C>0$ and $k\geq 3$, Algorithm~\ref{AlgParametrizedSetPacking} yields no better approximation guarantee than $\frac{k}{2}$.
\end{corollary}
\begin{proof}
	By Corollary~\ref{CorInstancesSetPacking}, we know that the approximation guarantee that Algorithm~\ref{AlgParametrizedSetPacking} achieves is no better than $\sup_{\epsilon >0} \frac{k}{2}-\epsilon = \frac{k}{2}$.
\end{proof}
\begin{proof}[Proof of Theorem~\ref{TheoNoBetterThanDMinus1Half}]
For $d\leq 3$, we have $\frac{d-1}{2}\leq 1$ and there is nothing to show. So assume that $d\geq 4$ and let $k:=d-1$. We first deal with the case $\alpha \leq 0$. To this end, consider a circuit $G$ on $2n\geq N_0$ vertices the weights of which alternate between $\frac{2}{d-1-\epsilon}\in(0,1)$ and $1$. Observe that all vertex degrees are $2<d-1$. Let $A$ consist of all vertices of weight $\frac{2}{d-1-\epsilon}$ and let $A^*$ comprise all vertices of weight $1$. Let $M$ be one of the two perfect matchings in $G$. As all weights are positive, each optimum solution to the MWIS picks precisely one vertex per edge of $M$.  Now, as $\left(\frac{2}{d-1-\epsilon}\right)^\alpha \geq 1^\alpha$ since $\alpha \leq 0$, $A$ is an independent set in $G$ of maximum $w^\alpha$-weight because it picks a maximum $w^\alpha$-weight vertex of each edge in $M$. In particular, there cannot be any local improvement improving $w^\alpha(A)$. On the other hand, $A^*$ is optimum for $w$, and $\frac{w(A^*)}{w(A)}=\frac{d-1-\epsilon}{2}$. Therefore, we can restrict ourselves to the case $\alpha >0$ in the following.\\ By \cite{ErdosSachs}, we know that for every $l\geq 4$, there exists a $k$-regular graph of girth at least $l$ such that its number $n(l)$ of vertices satisfies
\[
l\leq n(l)\leq 4\cdot\sum_{t=1}^{l-2} (k-1)^t = 4\cdot (k-1)\cdot \sum_{t=0}^{l-3}(k-1)^t = 4\cdot (k-1)\cdot \frac{(k-1)^{l-2}-1}{k-2}\leq 4\cdot (k-1)^{l-1}.
\]
In particular, $d\geq 4$ and, hence, $k-1\geq 2$ implies that $\log_{k-1} n(l)\leq l+1$.\\
 Let $\epsilon'_d:=1-(1-\frac{\epsilon}{d-1})^\alpha$ and let $0<\epsilon_d \leq \epsilon'_d$ with $\frac{1}{\epsilon_d}\in\mathbb{N}$. Pick $N_1$ such that for $n\geq N_1$, we have \begin{equation} \frac{\log(n)}{\log(k-1)}-1 > \frac{4}{\epsilon_d}\cdot \left(\log\left(\log\left(\frac{d+1}{2}\cdot n\right)\right)+\log(C)\right)\label{EqConditionN}.\end{equation} This is possible since the left hand side grows asymptotically faster than the right hand side. Pick $l\geq \max\{N_0,N_1\}$ and a $k$-regular graph $H$ of girth at least $l$ on $n$ vertices such that \[l\leq n\leq 4\cdot (k-1)^{l-1}.\] As we have seen before, this implies \begin{equation}
\frac{\log (|V(H)|)}{\log(k-1)}-1 = \frac{\log(n)}{\log(k-1)}-1\leq l.\label{EqLLog}
\end{equation} Define \[G:=(V(H)\cup E(H),\{\{v,e\}: v\in V(H), e\in E(H), v\in e\}\] and \[w:V(G)\rightarrow\mathbb{R}_{>0}, x\mapsto\begin{cases} 1 &, x\in V(H)\\ (1-\epsilon_d)^{\frac{1}{\alpha}} &,x\in E(H)\end{cases}.\] By $k=d-1$-regularity of $H$, every $v\in V(H)$ has degree $d-1 > 2$ in $G$, whereas each $e\in E(H)$ has degree $2 < d-1$ in $G$. In particular, all vertex degrees in $G$ are bounded from below by $2$ and from above by $d-1$, and $G$ is simple by construction. Moreover, by definition, $A:=V(H)$ constitutes a maximal independent set in $G$ and $A^*:=E(H)$ is an independent set in $G$ of cardinality $\frac{k}{2}\cdot |V(H)|=\frac{d-1}{2}\cdot |V(H)|$ (since $H$ is $k$-regular) and we get \[\frac{w(A^*)}{w(A)}=\dfrac{\frac{d-1}{2}\cdot (1-\epsilon_d)^\frac{1}{\alpha}\cdot |V(H)|}{|V(H)|}\geq \frac{d-1}{2}\cdot (1-\epsilon'_d)^\frac{1}{\alpha}=\left(1-\frac{\epsilon}{d-1}\right)\cdot\frac{d-1}{2}=\frac{d-1-\epsilon}{2}.\] Therefore, as $w(A^*)$ certainly defines a lower bound on the optimum value, it remains to show that there is not local improvement (w.r.t. $w^\alpha$) of size at most $C\cdot\log(|V(G)|)$ improving $A$. Assume towards a contradiction that $X\subseteq V(G)$ constitutes a local improvement of size at most $C\cdot\log(|V(G)|)$. We can assume without loss of generality that $X\subseteq E(H)$ because as $X$ is independent, we have $N(X\backslash V(H),X\cap V(H))=\emptyset$ and, therefore, \[N(X,V(H))= X\cap V(H)\dot{\cup} N(X\backslash V(H), V(H))\] and $X$ is a local improvement of $w^\alpha(A)=w^\alpha(V(H))$ if and only if $X\backslash V(H)$ is.\\
If $|X| < l$, then the subgraph $(N(X,V(H)), X)$ of $H$ is acyclic since the girth of $H$ is at least $l$. This implies that \mbox{$|N(X,V(H))|> |X|$}. As a consequence, given that $w^\alpha(x)< 1=w^\alpha(y)$ for all $x\in X\subseteq E(H)$, $y\in N(X,V(H))$, $X$ cannot be a local improvement. Hence, we must have $|X|\geq l$. As \[\frac{((1-\epsilon_d)^\frac{1}{\alpha})^\alpha}{1^\alpha}= 1-\epsilon_d,\] we must have $(1-\epsilon_d)\cdot |X| > |N(X,V(H))|$, i.e. \[|X|>\frac{|N(X,V(H))|}{1-\epsilon_d}= |N(X,V(H))|\cdot\left(1+\frac{\epsilon_d}{1-\epsilon_d}\right) \geq |N(X,V(H))|\cdot \left(1+\epsilon_d\right).\]  By Lemma $3.2$ from \cite{berman1994approximating} and since $\frac{1}{\epsilon_d}\in\mathbb{N}$, this implies that the subgraph $(N(X,V(H)), X)$ of $H$ possesses a cycle of length at most \begin{align*}\frac{4}{\epsilon_d}\cdot\log(|N(X,V(H))|)&\leq \frac{4}{\epsilon_d}\cdot \log(|X|)\leq \frac{4}{\epsilon_d}\cdot \log(C\cdot\log(|V(G)|))\\&=\frac{4}{\epsilon_d}\cdot \log(C\cdot\log(|V(H)|+|E(H)|))\\&=\frac{4}{\epsilon_d}\cdot \log\left(C\cdot\log\left(\frac{d+1}{2}\cdot |V(H)|\right)\right)\\&=\frac{4}{\epsilon_d}\cdot \left(\log(C)+\log\left(\log\left(\frac{d+1}{2}\cdot |V(H)|\right)\right)\right) \\&\stackrel{\eqref{EqConditionN}}{<} \frac{\log(|V(H)|)}{\log(k-1)}-1\stackrel{\eqref{EqLLog}}{\leq} l,\end{align*} a contradiction to the fact that the girth of $H$ is as least $l$. This finishes the proof.\end{proof}

%TODO local improvement algo not optimum for d=3?
\FloatBarrier

\section{Conclusion\label{SecConclusion}}
In this paper, we have seen how to use local search to approximate the weighted $k$-Set Packing Problem with an approximation ratio that gets arbitrarily close to $\frac{k}{2}$ as $k$ approaches infinity. At the cost of a quasi-polynomial running time, this result applies to the more general setting of the Maximum Weight Independent Set Problem in $d$-claw free graphs, yielding approximation ratios arbitrarily close to $\frac{d-1}{2}$. Moreover, we have seen that this result is asymptotically best possible in the sense that for no $\alpha\in\mathbb{R}$, a local improvement algorithm for the weighted $k$-Set Packing Problem that considers local improvements of $w^\alpha$ of logarithmically bounded size can produce an approximation guarantee better than $\frac{k}{2}$. As a consequence, our paper seems to conclude the story of (pure) local improvement algorithms for both the MWIS in $d$-claw free graphs and the weighted $k$-Set Packing Problem.\\ Hence, the search for new techniques beating the threshold of $\frac{d-1}{2}$, respectively $\frac{k}{2}$, might be one of the next goals for research in this area.  
%TODO: write some concluding remarks!!!
\bibliography{set_packing}

\appendix
\section{Proofs of Lemmata from the Analysis of SquareImp\label{AppendixAnalysisSquareImp}}
 \begin{proof}[Proof of Lemma~\ref{LemWeightsNeighborhoodsdminus1}]
 As $A^*$ is independent in $G$, we know that each $v\in V$ satisfies $|N(v,A^*)|\leq d-1$ because either $v\in A^*$ and $N(v,A^*)=\{v\}$, or $v\not\in A^*$ and $N(v,A^*)$ constitutes the set of talons of a claw centered at $v$, provided it is non-empty.
\end{proof}
\begin{proof}[Proof of Lemma~\ref{LemPropPositiveCharges}]
$\chrg{u}{v}>0$ implies $v=n(u)\in N(u,A)$ and, therefore, \begin{align*} w^2(N(u,A)\backslash\{v\})&=\sum_{x\in N(u,A)\backslash\{v\}}w^2(x)\\&\leq \sum_{x\in N(u,A)\backslash\{v\}} w(x)\cdot \max_{y\in N(u,A)} w(y)\\&=w(N(u,A)\backslash\{v\})\cdot w(v)\\&=(w(N(u,A))-w(v))\cdot w(v).\end{align*} From this, we get
 \begin{align*} 2\cdot\chrg{u}{v}\cdot w(v) &= (2\cdot w(u)-w(N(u,A)))\cdot w(v)\\ &= 2\cdot w(u)\cdot w(v)-w(N(u,A))\cdot w(v)\\ &\leq w^2(u)+w^2(v)-w(N(u,A))\cdot w(v)\\ &=w^2(u)-(w(N(u,A))-w(v))\cdot w(v)\\&\leq w^2(u)-w^2(N(u,A)\backslash\{v\})\end{align*} as claimed.
\end{proof}
\begin{proof}[Proof of Lemma~\ref{LemBoundCharges}] Assume for a contradiction that \[\sum_{u\in A^*: \chrg{u}{v}>0} \chrg{u}{v}>\frac{w(v)}{2} \] for some $v\in A$. Then $v\not\in A^*$ since \[\{u\in A^*:\chrg{u}{v}>0\}=\{v\}=N(v,A)=N(v,A^*) \] and \[\sum_{u\in A^*: \chrg{u}{v}>0}\chrg{u}{v}=\chrg{v}{v}=\frac{w(v)}{2} \] otherwise. Hence, $T:=\{u\in A^*: \chrg{u}{v}>0\}$ forms the set of talons of a claw centered at $v$. By Lemma~\ref{LemPropPositiveCharges}, it satisfies  \[w^2(T)=\sum_{u\in T} w^2(u)>\sum_{u\in T} w^2(N(u,A)\backslash \{v\}) +w^2(v)\geq w^2(N(T,A)), \] contradicting the fact that no claw improves $w^2(A)$.\end{proof}
\section{Conditions satisfied by our choices of \texorpdfstring{$\tilde{\epsilon}$}{epsilon tilde}, \texorpdfstring{$\epsilon'$}{epsilon prime} and \texorpdfstring{$d_{\delta}$}{d delta}\label{AppendixInequalities}}
\begin{equation}
0<\tilde{\epsilon}<\min\left\{2\delta, \frac{1}{2}\right\} \label{const0}
\end{equation}
\begin{equation}
0<\epsilon'\leq \frac{1}{20} \label{const1}
\end{equation}
\begin{equation}
\frac{1}{2}\cdot\left(1-\epsilon'-\frac{15}{4}\sqrt{2\epsilon'}\right)\cdot(1-\sqrt{2\epsilon'})\geq \frac{\epsilon'}{10} \label{const2}
\end{equation}
\begin{equation}
\frac{1-\sqrt{\epsilon'}}{2}\geq \frac{\epsilon'}{10}\label{const3}
\end{equation}
\begin{equation}
2+\epsilon'-\frac{1}{1-\sqrt{2\epsilon'}}\geq\sqrt{2\epsilon'}\label{const4}
\end{equation}
\begin{equation}
\frac{1}{2}< 1-\sqrt{2\epsilon'} \label{const7}
\end{equation}
\begin{equation}
\epsilon'\leq  (1-\sqrt{2\epsilon'})^2\label{const10}
\end{equation}
\begin{equation}
4\sqrt{\epsilon'}+\frac{4\sqrt{2\epsilon'}+8\epsilon'}{(1-\sqrt{2\epsilon'})^2}+\frac{18\epsilon'}{(1-\sqrt{2\epsilon'})^4} < \frac{\tilde{\epsilon}}{2}\label{const5}
\end{equation}
\begin{equation}
8\cdot\sqrt{\epsilon'}\leq \tilde{\epsilon} \label{const13}
\end{equation}
\begin{equation}
\forall d\geq d_{\delta}:\frac{20}{(d-1)\cdot \epsilon'}+\frac{\tilde{\epsilon}}{2}\leq\frac{\delta}{2}\label{const6}
\end{equation}
\begin{equation}
\forall d\geq d_{\delta}:1<\frac{d-1}{4}\cdot\tilde{\epsilon}\label{const8}
\end{equation}
\begin{equation}
\forall d\geq d_{\delta}:\frac{4\cdot (1-\sqrt{2\epsilon'})^{-2}}{d-5}<\frac{1}{2}\label{const9}
\end{equation}
\begin{equation}
\forall d\geq d_{\delta}: \frac{d-1}{d-5}< 2\label{const11}
\end{equation}
\begin{equation}
\forall d\geq d_{\delta}:9 < d\label{const12}
\end{equation}
\begin{proof}
	\eqref{const0}: Clear, as $\tilde{\epsilon}=\frac{\delta}{2}$ and $0<\delta<1$.\\
	\eqref{const1}: Also clear, as $\epsilon'=\frac{\delta^2}{2500}$.\\
	\eqref{const2}: We have
	\begin{align*}
	&\phantom{=}\frac{1}{2}\cdot \left(1-\epsilon'-\frac{15}{4}\sqrt{2\epsilon'}\right)\cdot(1-\sqrt{2\epsilon'}) \\&= \frac{1}{2}\cdot\left(1-\frac{\delta^2}{2500}-\frac{15}{4}\cdot\sqrt{2}\cdot\frac{\delta}{50}\right)\cdot\left(1-\sqrt{2}\cdot\frac{\delta}{50}\right)\\
	&> \frac{1}{2}\cdot\left(1-\frac{1}{2500}-\frac{15}{2}\cdot\frac{1}{50}\right)\cdot\left(1-\frac{1}{25}\right)\\
	&=\frac{1}{2}\cdot\frac{2124}{2500}\cdot \frac{24}{25}=\frac{25488}{62500}>\frac{1}{25000} \geq \frac{\epsilon'}{10}.
	\end{align*}
	\eqref{const3}: 
	\[\frac{1-\sqrt{\epsilon'}}{2}\geq \dfrac{1-\frac{1}{50}}{2} >\frac{1}{4} > \frac{\epsilon'}{10}\]
	\eqref{const4}:	\begin{align*}
	2+\epsilon'-\frac{1}{1-\sqrt{2\epsilon'}}&\geq 2 -\dfrac{1}{1-\sqrt{\frac{4}{2500}}}= 2-\dfrac{1}{1-\frac{1}{25}}\\
	&=2-\frac{25}{24} = \frac{23}{24} > \frac{1}{25}=\sqrt{\frac{4}{2500}}> \sqrt{2\epsilon'}
	\end{align*}
	\eqref{const7}: \begin{align*}
	\frac{1}{2}< 1 - \frac{1}{25}= 1-\sqrt{\frac{4}{2500}} < 1-\sqrt{2\epsilon'} 
	\end{align*}
	\eqref{const10}:
	\[\epsilon'\leq \frac{1}{2500} < \frac{576}{625} =\left(1-\frac{1}{25}\right)^2=\left(1-\sqrt{\frac{4}{2500}}\right)^2\leq (1-\sqrt{2\epsilon'})^2\]
	\eqref{const5}:
	\begin{align*}
	\phantom{=}&4\sqrt{\epsilon'}+\frac{4\sqrt{2\epsilon'}+8\epsilon'}{(1-\sqrt{2\epsilon'})^2}+\frac{18\epsilon'}{(1-\sqrt{2\epsilon'})^4}\\
	&=4\sqrt{\frac{\delta^2}{2500}}+\dfrac{4\sqrt{2\cdot\frac{\delta^2}{2500}}+8\cdot\frac{\delta^2}{2500}}{\left(1-\sqrt{2\cdot\frac{\delta^2}{2500}}\right)^2}+\dfrac{18\cdot\frac{\delta^2}{2500}}{\left(1-\sqrt{2\cdot\frac{\delta^2}{2500}}\right)^4}\\
	&\leq \frac{2}{25}\cdot\delta + \dfrac{4\cdot\frac{3}{2}\cdot\frac{\delta}{50}+ \frac{2}{625}\cdot\delta^2}{\left(1-\frac{3}{2}\cdot\frac{\delta}{50}\right)^2}+\dfrac{\frac{9}{1250}\cdot\delta^2}{\left(1-\frac{3}{2}\cdot\frac{\delta}{50}\right)^4} \\
	&\leq \frac{2}{25}\cdot\delta + \dfrac{\frac{3}{25}\cdot \delta+ \frac{2}{625}\cdot\delta^2}{\left(1-\frac{3}{100}\right)^2}+\dfrac{\frac{9}{1250}\cdot\delta^2}{\left(1-\frac{3}{100}\right)^4} \\
	&\leq \frac{2}{25}\cdot\delta + \dfrac{\frac{3}{25}\cdot \delta+ \frac{2}{625}\cdot\delta^2}{\frac{9409}{10,000}}+\dfrac{\frac{9}{1250}\cdot\delta^2}{\frac{88,529,281}{100,000,000}} \\
	&\leq \frac{2}{25}\cdot\delta + \frac{1200}{9409}\cdot\delta + \frac{32}{9409}\cdot\delta^2+\frac{720,000}{88,529,281}\cdot\delta^2\\
	&\leq \left(\frac{2}{25}+\frac{1232}{9409}+\frac{720,000}{88,529,281}\right)\cdot\delta \\
	&<(0.08+0.131+0.01)\cdot \delta = 0.221\cdot\delta < \frac{\delta}{4}=\frac{\tilde{\epsilon}}{2} \end{align*} 
	\eqref{const13}:
	\[8\cdot\sqrt{\epsilon'}=\frac{8}{50}\cdot\delta < \frac{\delta}{2} = \tilde{\epsilon}\]
	\eqref{const6}: For $d\geq d_{\delta}=\frac{200,000}{\delta^3}+1$, we get
	
	\[ \frac{20}{(d-1)\cdot \epsilon'}+\frac{\tilde{\epsilon}}{2} \leq  \dfrac{20}{\frac{\delta^2}{2500}\cdot\frac{200,000}{\delta^3}}+\frac{\delta}{4}=\frac{\delta}{4}+\frac{\delta}{4}=\frac{\delta}{2}\] 
	\eqref{const8}: For $d\geq d_{\delta}=\frac{200,000}{\delta^3}+1$, we get 
	\[\frac{d-1}{4}\cdot\tilde{\epsilon}=\frac{200,000}{4\cdot\delta^3}\cdot \frac{\delta}{2} = \frac{25,000}{\delta^2}\geq 25,000 > 1.\]
	\eqref{const9}: For $d\geq d_{\delta}=\frac{200,000}{\delta^3}+1$, we get 
	\begin{align*}
	\frac{4\cdot (1-\sqrt{2\epsilon'})^{-2}}{d-5} &=\dfrac{4}{\left(1-\sqrt{2\cdot\frac{\delta^2}{2500}}\right)^2\cdot(\frac{200,000}{\delta^3}-4)}\\
	&\leq \dfrac{4}{\left(1-\frac{1}{25}\right)^2\cdot 199,996} = \frac{2500}{115,197,696}<\frac{1}{2}
	\end{align*}
	\eqref{const11} and \eqref{const12}: $\frac{d-1}{d-5} < 2$ is, for $d > 5$, equivalent to $d-1 < 2\cdot(d-5)\Leftrightarrow 9 < d$. By our choice of $d_\delta$ and since $0<\delta<1$, the latter is clearly true.
\end{proof}
%TODO also works for set packing
\end{document}